\documentclass[article, nojss]{jss}

\usepackage{caption}
\usepackage{thumbpdf,lmodern}
\usepackage{framed}
\usepackage{bm}
\usepackage{natbib}
\usepackage{color}
\usepackage[plain,noend]{algorithm2e}
\usepackage{graphicx}
\usepackage{amsmath,amssymb}
\usepackage[utf8]{inputenc}
\usepackage{multirow}
\usepackage{booktabs}
\usepackage{framed}

\usepackage[usenames,dvipsnames]{xcolor}
\newcommand{\comment}[1]{#1}  
\newcommand{\commentred}[1]{#1}  
\newcommand{\revised}[1]{#1}

\newcommand{\yhat}{\hat{y}}
\newcommand{\yS}{S}
\newcommand{\QQ}{\comment{\mathbf{Q}}} 
\newcommand{\aphi}{\comment{a_\phi}} 
\newcommand{\bphi}{\comment{b_\phi}} 
\newcommand{\Bsv}{\comment{B_\sigma}} 
\newcommand{\OOmega}{\comment{\Omegav}} 
\newcommand{\SSigma}{\comment{\Sigmam}} 
\newcommand{\bct}[2]{\comment{\beta_{{#1}{#2}}}} 

\newcommand{\xv}{\comment{\bm x}}
\newcommand{\cv}{\comment{\bm c}}
\newcommand{\imm}[1]{^{\comment{(#1)}}}

\newcommand{\weg}[1]{}

\newcommand{\cm}{{\mathbf c}}
\newcommand{\Fm}{{\mathbf F}}

\renewcommand{\vfill}{\vspace*{\fill}}

\newcommand{\bfz}{{\mathbf{0}}}

\newcommand{\Omegav}{\boldsymbol \Omega}

\newcommand{\betac}{\beta} 
\newcommand{\betar}{\boldsymbol{\betac}} 
\newcommand{\betav}{\betar} 

\newcommand{\Normult}[2]{ \mathcal{N} _{#1}\left(#2\right)}

\newcommand{\yc}{y}
\newcommand{\ym}{{\mathbf \yc}} 

\newcommand{\pdf}[3]{f_{ {\footnotesize #1}}(#2;#3)}
\newcommand{\Normalpdfa}[2]{\pdf{\mathcal{N}}{#1}{#2}}
\newcommand{\Normal}[1]{ \mathcal{N}\left(#1\right)}

\newcommand{\Gammad}[1]{ \mathcal{G}\left(#1\right)}

\newcommand{\Sigmam}{\boldsymbol{\Sigma}} 

\newcommand{\Diag}[1]{\mbox{\rm Diag}\left(#1\right)}

\newcommand{\Gammainv}[1]{\mathcal{G}^{-1} \left(#1\right)}

\newcommand{\Ew}[1]{\mbox{\rm E}(#1)}   

\newtheorem{alg}{Algorithm}

\Plainauthor{Peter Knaus, Angela Bitto-Nemling, Annalisa Cadonna, Sylvia Fr\"uhwirth-Schnatter}

\author{Peter Knaus\\WU Vienna
\And Angela Bitto-Nemling
\And Annalisa Cadonna\\AI CoE, Crayon Austria
\AND Sylvia Fr\"uhwirth-Schnatter\\WU Vienna }

\title{Shrinkage in the Time-Varying Parameter Model Framework Using the \proglang{R} Package \pkg{shrinkTVP}}
\Plaintitle{Shrinkage in the Time-Varying Parameter Model Framework Using the R Package \pkg{shrinkTVP}}
\Shorttitle{Shrinkage for TVP Models Using \pkg{shrinkTVP}}

\Abstract{
Time-varying parameter (TVP) models are widely used in time series analysis to
flexibly deal with processes which gradually change over time. However, the risk of
overfitting in TVP models is well known. This issue can be dealt with using appropriate
global-local shrinkage priors, which pull time-varying parameters towards static ones. In this paper, we introduce the  \proglang{R} package \pkg{shrinkTVP} \citep{kna-etal:shr},  which
provides a fully Bayesian implementation of shrinkage priors for TVP models, taking advantage of recent developments in the literature, in particular those of \cite{bit-fru:ach} \revised{and \cite{cad-etal:tri}}.
The package \pkg{shrinkTVP}  allows for posterior simulation of the parameters through an efficient Markov Chain Monte Carlo
scheme. Moreover, summary and visualization methods, as well as the possibility of assessing predictive performance through log-predictive density scores, are provided.
The computationally intensive tasks have been implemented in \proglang{C++} and interfaced with \proglang{R}.  The paper includes a brief overview
of the models and shrinkage priors implemented in the package. Furthermore, core functionalities are illustrated, both with simulated and real data.
}


\Address{
Peter Knaus\\
Institute for Statistics and Mathematics\\
Department of Finance, Accounting and Statistics\\
WU Vienna University of Economics and Business\\
Welthandelsplatz 1, Building D4, Entrance A, 3rd Floor\\
1020 Vienna, Austria\\
E-mail: \email{peter.knaus@wu.ac.at}
URL: \url{https://www.wu.ac.at/statmath/faculty-staff/faculty/peter-knaus}
}

\begin{document}


\section[Introduction]{Introduction} \label{sec:intro}

Time-varying parameter (TVP) models are
widely used in time series analysis, because of their flexibility and ability to capture gradual changes in the model \comment{parameters} 
over time. The popularity of TVP models in macroeconomics and finance is based on the fact that, in most applications, the influence of certain predictors on the outcome variables varies over time \citep{pri:tim, dan-hal:pre, bel-etal:hie_tv}. TVP models, while capable of reproducing salient features of the data in a very effective way, present a concrete risk of overfitting, as often only a small subset of the parameters are time-varying. Hence, in the last decade, there has been a growing need for models and methods able to discriminate between time-varying and static parameters in TVP models.
A key contribution in this direction was the introduction of the non-centered parameterization of TVP models in  \cite{fru-wag:sto}, which recasts the problem of variance selection and shrinkage in terms of variable selection, thus allowing any tool used to this end in multiple regression models to be used to perform selection or shrinkage of variances.
\cite{fru-wag:sto} employ a spike and slab prior, while continuous shrinkage priors have been utilised as a regularization alternative in, e.g., \cite{bel-etal:hie_tv},
\cite{bit-fru:ach} and \cite{cad-etal:tri}. For an excellent review of shrinkage priors, with a particular focus on high dimensional regression, the reader is directed to  \cite{bha-etal:las}.

In this paper, we describe the \proglang{R}  package  \pkg{shrinkTVP}  \citep{kna-etal:shr} for Bayesian TVP models with shrinkage. The package is available under the general public license (GPL $\geq$ 2) from the Comprehensive R
Archive Network (CRAN) at \url{https://cran.r-project.org/web/packages/shrinkTVP}.
The package efficiently implements recent developments in the Bayesian literature, in particular the ones
presented in \cite{bit-fru:ach} and \cite{cad-etal:tri}. The  computationally intensive Markov chain Monte Carlo (MCMC) algorithms
in the package are written in \proglang{C++} and interfaced with \proglang{R} \citep{R}  via the \pkg{Rcpp} \citep{edd-bal:ext} and the \pkg{RcppArmadillo} \citep{edd-san:acc} packages.
This approach combines the ease-of-use of R and its underlying functional programming paradigm with the computational speed of \proglang{C++}.

\commentred{The  package \pkg{shrinkTVP} is designed to provide an easy entry point for fitting TVP models with shrinkage priors, while also giving more experienced users the option to adapt the model to their needs. This is achieved by providing a robust baseline model that can be estimated by only passing the data, while also allowing the user to specify more advanced options.
Additionally, the \pkg{shrinkTVP} package is  designed to
ensure compatibility with well-known times series formats and to complement other packages. As input objects,  time series from the  \proglang{R} packages  \pkg{zoo}  \citep{zei:zoo} and \pkg{xts} \citep{xts} as well as time series formats like \code{ts} are supported.
Estimation output   is compatible  with the popular  \proglang{R} package  \pkg{coda}  \citep{plu:cod}   which can  be easily applied for  convergence diagnostic tests, among others.
Coupled with intuitive summary and plot methods, \pkg{shrinkTVP} is  a package that is  easy to use while remaining highly flexible. }

\commentred{\pkg{shrinkTVP} is, to our knowledge, the only \proglang{R} package that combines TVP models with shrinkage priors on the time-varying components  in a Bayesian framework.
Several  \proglang{R} packages deal  with statistical inference for  various specific classes of  state space models, of which TVP models are a special case.
The most popular \proglang{R} package in this field is \pkg{dlm} \citep{pet:anr},  a comprehensive package providing routines for maximum likelihood estimation, Kalman filtering and smoothing, and Bayesian analysis for dynamic linear models  (DLMs).
The  accompanying  book \citep{pet-etal:dyn} introduces the methodology and many \proglang{R} code examples.
As of now, priors are not designed to encourage shrinkage and   \pkg{shrinkTVP}  complements   \pkg{dlm} in this regard.}

\commentred{The \proglang{R} package \pkg{bvarsv} \citep{kru:bva} implements  Bayesian inference for  vector autoregressive (VAR) models with time-varying parameters  (TVP-VAR)
and stochastic volatility for multivariate time series  as introduced by  \citep{pri:tim}. 
 We refer to \citep{del-pro:tim2} for details on  the MCMC algorithm  and a later correction of the original scheme.
 In addition to the very user friendly estimation function \code{bvar.sv.tvp},  \pkg{bvarsv} provides posterior predictive distributions and  enables impulse response analysis.  The package  includes the macroeconomic data set analysed  in \citep{pri:tim}  as   example data set,  \code{usmacro.update},  which we use in our predictive exercise in Section~\ref{sec:usmacro} to showcast the effect of introducing shrinkage priors on time-varying parameters.}

\commentred{Additional packages emerged very recently. 
The \proglang{R} package \pkg{tvReg} \citep{tvreg}  presents a user friendly compendium of  many common linear  TVP models,  including  standard linear regression as well as autoregressive,  seemingly unrelated equation  and  VAR models.
Estimation is based on kernel smoothing techniques. 
For an illustrative  application,  a TVP-VAR(4) model  is fitted  
  to the \code{usmacro.update} data set mentioned above,   using the   function  \code{tvVAR}.}
  \commentred{The  \proglang{R} package \pkg{walker} \citep{walker}  facilitates the estimation of DLMs and generalized  DLMs using MCMC  algorithms provided by \proglang{Stan} \citep{carpenter2017stan}.  For inference, the importance sampling method of \citep{ISMCMC} is implemented within a Hamiltonian Monte Carlo framework.}
\commentred{The \proglang{R} package \pkg{bsts} \citep{sco:bst} performs Bayesian analysis for  structural time series models, a highly  relevant  class of  state space models   including DLMs.  \pkg{bsts} is a very powerful package that  allows
shrinkage  for static  regression coefficients using spike and slab priors.
However, as for any other  packages mentioned above, variation of the dynamic  components is not regularized in   \pkg{bsts}.}

 \commentred{A main contribution of  the  package \pkg{shrinkTVP}  is  bridging the  active field of  \proglang{R} packages for state space models with the even more active field of \proglang{R} packages that  provide regularization and shrinkage methods  for common regression type models.

Among others,  \pkg{ncvreg} \citep{ncvreg} is useful for fitting standard penalized regression estimators,
 \pkg{glmnet} \citep{sim-etal:reg} allows  elastic-net regularization for a variety of models,
   \pkg{horseshoe} \citep{pas:hor}  implements the  horseshoe prior, while  \pkg{shrink} \citep{dun-etal:shr} provides various shrinkage methods for linear, generalized linear, and Cox regression models. \pkg{biglasso} \citep{zen:big} aims at very fast lasso-type regularization  for high-dimensional linear regression.
Recent \proglang{R} packages include   \pkg{NormalBetaPrime} \citep{bai:nor}
 for  Bayesian univariate
and \pkg{MBSP} \citep{MBSP} for  Bayesian multivariate linear regression analysis using, respectively,  the normal-beta prime  and the three parameter beta normal family for inducing shrinkage. The \proglang{R} package  \pkg{monomvn} \citep{monomvn}  employs a  normal-gamma prior in the specific situation
of  Bayesian inference for  multivariate normal and Student-$t$ data with a monotone  pattern of missing data.
}

 \commentred{The remainder of the paper is organized as follows.
Section~\ref{sec:model} briefly introduces TVP \comment{models} and normal-gamma-gamma shrinkage priors, and describes the MCMC algorithms for posterior simulation. The package \pkg{shrinkTVP} is introduced in Section~\ref{sec:pkgshrinkTVP}. In particular, we illustrate how to run the MCMC sampler using the main function \code{shrinkTVP}, how to choose a specific model, and how to conduct posterior inference using the return object of \code{shrinkTVP}. Section~\ref{sec:LPDS} explains how to assess model   performance   by calculating log-predictive density scores (LPDSs), and how to use LPDSs to compare the predictive performances of different priors. This is illustrated using the \code{usmacro.update} data set. 
Finally, Section~\ref{sec:conclusions} concludes the paper.}

\section{Model specification and estimation } \label{sec:model}

\subsection{TVP models}

Let us recall the state space form of a TVP model. For $t = 1, \ldots, T$, we have that
\begin{equation}
\label{eq:centeredpar}
\begin{aligned}
&y_{t} =   \bm x_t \bm {\beta_{t}}  +  \epsilon_{t} , \qquad
\epsilon_{t} \sim \mathcal N (0,\sigma^2_t), \\
& \bm {\beta}_{t} =  \bm {\beta}_{t-1} + \bm w_{t}, \qquad   \bm w_{t}  \sim \mathcal N_d (0, \QQ),
\end{aligned}
\end{equation}
where $y_t$ is a univariate response variable and $\bm x_t = (x_{t 1}, x_{t 2}, \ldots, x_{t  d})$ is a $d$-dimensional row vector containing the regressors at time $t$, with $x_{t 1}$ corresponding to the intercept.
For simplicity, we assume here that $\QQ = \text{Diag}(\theta_1, \ldots, \theta_d)$ \comment{is a diagonal matrix}, implying that the state innovations are conditionally independent.
Moreover, we assume the initial value follows a normal distribution, i.e., $\bm \beta_{0} \sim \mathcal N_d (\bm \beta, \QQ)$\comment{, with initial mean
$\bm \beta = (\beta_1, \ldots, \beta_d)$}.
Model (\ref{eq:centeredpar}) can be rewritten equivalently in the non-centered parametrization as
\begin{equation}
\label{eq:noncenteredpar}
\begin{aligned}
&y_t= \bm x_t \bm \beta +   \bm x_t \text{Diag}(\sqrt{\theta}_1, \ldots, \sqrt{\theta}_d)
\tilde{\bm \beta}_{t} +  \epsilon_t, \quad  \epsilon_t \sim \mathcal N (0,\sigma^2_t),\\
&\tilde{\bm \beta}_{t} =\tilde {\bm \beta}_{t-1} + \tilde{\bm u}_{t}, \qquad \tilde{\bm u}_{t} \sim  \mathcal N_d (0, I_d),
\end{aligned}
\end{equation}
with $ \tilde{\bm \beta}_{0} \sim \mathcal{N}_d (\bm 0, I_d)$, where $I_d$ is the $d$-dimensional identity matrix.

\pkg{shrinkTVP} is capable of modelling the observation error both homoscedastically, i.e., $\sigma^2_t \equiv \sigma^2$ for all $t = 1, \ldots, T$ and heteroscedastically, via a stochastic volatility \comment{(SV)}  specification. In the latter case, the log-volatility $h_t = \log \sigma^2_t$ follows an AR(1) model \citep{jac-etal:bayJBES,kas-fru:anc, kas:dea}. More specifically,
\begin{eqnarray}
	\label{eq:svht}
	h_t | h_{t-1}, \mu, \phi, \sigma_\eta^2 \sim \mathcal{N} \left(\mu +\phi ( h_{t-1}-\mu),\sigma^2_\eta \right),
\end{eqnarray}
with initial state $h_0 \sim \mathcal N \left(\mu, \sigma_\eta^2/(1-\phi^2) \right)$.
The stochastic volatility model on the errors can prevent the detection of spurious variations in the TVP coefficients \citep{nak:tim, sim:com} by capturing some of the variability in the error term.

\subsection{Prior Specification} \label{sec:priors}

\subsubsection{Shrinkage priors on variances and model parameters}

We place conditionally independent \revised{normal-gamma-gamma (NGG) priors \citep{cad-etal:tri,gri-bro:hie}}, both on the standard deviations of the innovations, that is the $\sqrt{\theta_j}$'s,
and on the means of the initial value $\beta_j$, for $j = 1, \ldots, d$. Note that, in the case of the standard deviations,
this can equivalently be seen as a triple gamma prior on the innovation variances $\theta_j$, for $j = 1, \ldots, d$.
\revised{The NGG can be represented as a conditionally normal distribution, where the component specific variance is itself a compound probability distribution resulting from two gamma distributions. In this representation, it looks as follows
\begin{eqnarray} 
\sqrt{\theta}_j|\xi^{2}_j  \sim \Normal{0,\xi_j^{2}}, \qquad \xi_j^2|a^\xi,\kappa_j^2  \sim  \Gammad{a^\xi,\frac{a^\xi \kappa_j^2}{2}}, \quad
\kappa_j^2| c^\xi, \kappa^2_B \sim \Gammad{c^\xi, \frac{c^\xi}{\kappa^2_B}} \label{eq:normalTG1}\\
\beta_{j}|\tau^2_j \sim  \Normal{0,\tau^2_j}, \qquad \tau_j^2|a^\tau ,\lambda_j^2 \sim  \Gammad{a^\tau,\frac{a^\tau \lambda_j^2}{2}} \quad \lambda_j^2| c^\tau, \lambda^2_B \sim \Gammad{c^\tau, \frac{c^\tau}{\lambda^2_B}}. \label{eq:normalTG2}
\end{eqnarray}
Letting $c^\xi$ and $c^\tau$ go to infinity results in a normal-gamma (NG) prior \citep{gri-bro:inf} on the $\sqrt{\theta}_j$'s and $\beta_j$'s. It has a representation as a conditionally normal distribution, with the component specific variance following a gamma distribution, that is
\begin{eqnarray} 
	\sqrt{\theta}_j|\xi^{2}_j  \sim \Normal{0,\xi_j^{2}}, \qquad \xi_j^2|a^\xi,\kappa_B^2  \sim  \Gammad{a^\xi,\frac{a^\xi \kappa_B^2}{2}}, \label{eq:normalDG1}\\
	\beta_{j}|\tau^2_j \sim  \Normal{0,\tau^2_j}, \qquad \tau_j^2|a^\tau ,\lambda_B^2 \sim  \Gammad{a^\tau,\frac{a^\tau \lambda_B^2}{2}}. \label{eq:normalDG2}
	\end{eqnarray}
From here, letting $a^\xi$ and $a^\tau$ go to infinity yields a normal prior with fixed variance, also known as ridge regression:
\begin{eqnarray} 
	\sqrt{\theta}_j|\kappa_B^2  \sim \Normal{0,\frac{2}{\kappa^2_B}}, \label{eq:normalRidge1}\\
	\beta_{j}|\lambda_B^2 \sim  \Normal{0,\frac{2}{\lambda_B^2}}. \label{eq:normalRidge2}
	\end{eqnarray}
	We refer to $a^\xi$ and $a^\tau$ as the pole parameters, as marginally more mass is placed around zero as they become smaller. $c^\xi$ and $c^\tau$ are referred to as the tail parameters, as they control the amount of mass in the tails of the distribution, with smaller values equating to heavier tails. Finally, the parameters $\kappa_B^2$ and $\lambda_B^2$ are dubbed the {global shrinkage parameters}, as they influence how strongly {all} parameters are pulled to zero. The larger $\kappa_B^2$ and $\lambda_B^2$, the stronger  \comment{this effect}.}

	\revised{One of the key benefits of the NGG prior is that many interesting shrinkage priors are contained within it as special or limiting cases. Beyond the NG prior mentioned above, two such cases are the horseshoe prior \citep{car-etal:han} and the Bayesian Lasso \citep{par-cas:bay}. The former results from an NGG prior with the pole and tail parameters equal to $0.5$, while the latter is a special case of the NG prior with a pole parameter fixed to one. As the connection between the NGG prior and the horseshoe prior may not be entirely obvious from the parameterization presented here, the interested reader is referred to \cite{cad-etal:tri} for details.}

The parameters $a^\xi$, $a^\tau$, $c^\xi$, $c^\tau$, $\kappa_B^2$ and $\lambda_B^2$ can be learned from the data through appropriate prior distributions. Results from \cite{cad-etal:tri} motivate the use of different distributions for these parameters under the NGG and NG prior. In the NGG case, the scaled global shrinkage parameters conditionally follow F distributions, depending on their respective pole and tail parameters:
\begin{align} \label{eq:NGG_glob_pri}
	\frac{\kappa_B^2}{2}|a^\xi, c^\xi \sim F (2a^\xi , 2c^\xi), \qquad \frac{\lambda_B^2}{2}|a^\tau, c^\tau \sim F (2a^\tau , 2c^\tau).
\end{align}
The scaled tail and pole parameters, in turn, follow beta distributions:
\begin{eqnarray} \label{eq:NGG_tail_pole_pri}
	2a^\xi \sim \mathcal{B}\left(\alpha_{a^\xi}, \beta_{a^\xi}\right), \qquad 2c^\xi \sim \mathcal{B}\left(\alpha_{c^\xi}, \beta_{c^\xi}\right), \\
	2a^\tau \sim \mathcal{B}\left(\alpha_{a^\tau}, \beta_{a^\tau}\right), \qquad 2c^\tau \sim \mathcal{B}\left(\alpha_{c^\tau}, \beta_{c^\tau}\right).
\end{eqnarray}
These priors are chosen as they imply a uniform prior on a suitably defined model size, see \cite{cad-etal:tri} for details.
In the NG case the global shrinkage parameters follow independent gamma distributions:
\begin{align} \label{eq:NG_glob_pri}
	\kappa_B^2 \sim \mathcal G (d_1, d_2), \qquad \lambda_B^2 \sim \mathcal G (e_1, e_2).
\end{align}
In order to learn the pole parameters in the NG case, we generalize the approach taken in \cite{bit-fru:ach} and place the following gamma distributions as priors:
\begin{align} \label{eq:equNG02}
	a^\xi\sim \mathcal G(\alpha_{a^\xi} , \alpha_{a^\xi}\beta_{a^\xi}), \qquad
	a^\tau \sim \mathcal G(\alpha_{a^\tau} , \alpha_{a^\tau}\beta_{a^\tau}),
\end{align}
which correspond to the exponential priors used in \cite{bit-fru:ach} when \revised{$\alpha_{a^\xi}=1$ and $\alpha_{a^\tau}=1$. The parameters $\alpha_{a^\xi}$ and $\alpha_{a^\tau}$} act as degrees of freedom and allow the prior to be bounded away from zero.

\subsubsection{Prior on the volatility parameter}

In the homoscedastic case we employ a hierarchical prior, \comment{where the scale of an inverse gamma prior  for $\sigma^2$ follows a gamma distribution}, that is,
\begin{eqnarray} \label{eq:priorsigma}
	\sigma^2|C_0 \sim \Gammainv{c_0,C_0}, \qquad  C_0 \sim \Gammad{g_0,G_0},
\end{eqnarray}
with hyperparameters $c_0$, $g_0$, and $G_0$.

In the case of stochastic volatility, the priors on the parameters $\mu$, $\phi$ and $\sigma^2_\eta$ in Equation~\eqref{eq:svht} are chosen as in \citet{kas-fru:anc}, that is
\begin{eqnarray} \label{eq:volpriors}
	\mu  \sim \mathcal{N}( b_\mu, B_\mu ), \quad \dfrac{\phi +1 }{2} \sim \mathcal{B}(\aphi, \bphi), \quad  \sigma^2_\eta \sim \mathcal{G}(1/2, 1/2 \Bsv ),
\end{eqnarray}
\comment{with hyperparameters $b_\mu, B_\mu, \aphi, \bphi,$ and $\Bsv$.}

\subsection{MCMC sampling algorithm}

\label{sec:MCMC}
The package \pkg{shrinkTVP} implements an MCMC Gibbs sampling algorithm with Metropolis-Hastings steps to obtain draws from the posterior distribution of the model parameters. 
Here, we roughly sketch the sampling algorithm and refer the interested reader to \cite{bit-fru:ach} \revised{and \cite{cad-etal:tri}} for further details.

\newpage 

\commentred{\begin{alg} \label{facsvalg}
\mbox{\rm Gibbs Sampling Algorithm}
 \begin{enumerate} \itemsep 0mm
\item[\mbox{\rm 1.}] \mbox{\rm Sample the latent states} $\tilde \betav =( \tilde \betav_0, \ldots, \tilde \betav_T)$ \mbox{\rm  in the non-centered parametrization from  a}\\
   \mbox{\rm  multivariate  normal  distribution;}
\item[\mbox{\rm 2.}] \mbox{\rm Sample jointly} $\beta_1, \dots,\beta_d,$ \mbox{\rm  and} $\sqrt{\theta_1},\dots,\sqrt{\theta_d}$ \mbox{\rm  in the non-centered parametrization from}\\
    \mbox{\rm  a multivariate  normal  distribution;}
\item[\mbox{\rm 3.}] \mbox{\rm Perform an ancillarity-sufficiency interweaving step and redraw  each} $\beta_1, \dots,\beta_d$ \mbox{\rm from a}\\
\mbox{\rm   normal  distribution and each} ${\theta_1},\dots,{\theta_d}$ \mbox{\rm from a generalized inverse Gaussian distribution}\\
\mbox{\rm  using \pkg{GIGrvg} \citep{hoe-ley:gig};}
\item[\mbox{\rm 4.}] \revised{ \mbox{\rm Sample the prior variances} $\xi^2_1, \dots \xi^2_d$ \mbox{\rm  and}  $\tau^2_1, \dots \tau^2_d$ \mbox{\rm and the component specific hyper-} \\
\mbox{\rm parameters. Sample (where required) the pole, tail and global shrinkage parameters.} \\
\mbox{\rm In the NGG case, this is done by emplyoing steps (c) - (f) from Algorithm 1 in} \\
\mbox{\rm \cite{cad-etal:tri}. In the NG case  steps (d) and (e) from Algorithm 1 in } \\
\mbox{\rm \cite{bit-fru:ach} are used. In the ridge regression case simply}\\ 
\mbox{\rm set} $\xi^2_j = 2/\kappa^2_B$ \mbox{\rm and} $\tau^2_j = 2/\lambda^2_B$, \mbox{\rm for} $d = 1, \dots, d$.
}
\item[\mbox{\rm 5.}]  \mbox{\rm Sample the error variance} $\sigma^2$ \mbox{\rm from an inverse gamma distribution in the homoscedastic}
 \mbox{\rm case or, in the SV case, sample the level} $\mu$, \mbox{\rm the persistence} $\phi$, 
    \mbox{\rm  the volatility of the vola-}\\ \mbox{\rm tility} $\sigma^2_{\eta}$  \mbox{\rm and the log-volatilities} $\bm h= (h_0, \ldots, h_T)$
     \mbox{\rm  using \pkg{stochvol} \citep{kas:dea}.}
\end{enumerate}
\end{alg}}

\revised{Step 4 presents a fork in the algorithm, as different parameterizations are used in the NGG and NG case, as to improve mixing. For details on the exact parameterization used in the NGG case, see \cite{cad-etal:tri}. Additionally, not all sampling steps are performed in all prior setups. If, for example, the user has defined that $\kappa_B^2$ should not be learned from the data, then this step is not executed.}

One key feature of the algorithm is the joint sampling of the time-varying parameters $\tilde{\bm \beta}_t$, for $t=0, \ldots, T$ in step 1 of Algorithm~\ref{facsvalg}. We employ the procedure described in
\cite{mcc-etal:sim} which exploits the sparse, block tri-diagonal structure of the precision matrix of the full conditional distribution of
 \comment{$\tilde \betav =( \tilde \betav_0, \ldots, \tilde \betav_T)$},  
 to speed up computations.

Moreover, as described in \cite{bit-fru:ach}, in step 3 we make use of the  ancillarity-sufficiency interweaving strategy (ASIS) introduced by  \citet{yu-men:cen}. ASIS is well known to improve mixing by sampling certain parameters both in the centered and non-centered parameterization.
This strategy has been successfully applied to univariate SV models \citep{kas-fru:anc}, multivariate factor SV models \citep{kas-etal:eff}  and  dynamic linear state space models \citep{sim-etal:int}.

\revised{
\paragraph{Adaptive Metropolis-within-Gibbs}
For the pole and tail parameters, no full conditionals exist and a Metropolis-Hastings step has to be performed. To improve mixing, \code{shrinkTVP} supports adaptive Metropolis-within-Gibbs as in \cite{rob-ros:exa}. The algorithm works as follows. For each parameter $i$ that is being learned from the data, let $s_i$ represent the standard deviation of the proposal distribution. After the $n^{th}_i$ batch of $m_i$ iterations, update $s_i$ according to the following rule:
\begin{itemize}
	\itemsep -1mm
	\item increase the log of $s_i$ by $\text{min}(c_i, n_i^{1/2})$ if the acceptance rate of the previous batch was above $d_i$ or
	\item decrease the log of $s_i$ by $\text{min}(c_i, n_i^{1/2})$ if the acceptance rate of the previous batch was below $d_i$.
\end{itemize}
The starting value of $s_i$, $m_i$, $c_i$ and $d_i$ can all be set by the user. Additionally, if adaptive Metropolis-within-Gibbs is not desired, it can be switched off and a simple Metropolis-Hastings step will be performed.
}

\section[The shrinkTVP package]{The \pkg{shrinkTVP} package}
\label{sec:pkgshrinkTVP}

\subsection{Running the model}
The core function of the package \pkg{shrinkTVP} is the function \code{shrinkTVP}, which serves as an R-wrapper for the actual sampler coded in \proglang{C++}. The function works out-of-the-box, meaning that estimation can be performed with minimal user input. With default settings, the TVP model in \comment{Equation}~\eqref{eq:centeredpar} is estimated in a Bayesian fashion with \revised{the NG prior defined in equations~\eqref{eq:normalDG1}, \eqref{eq:normalDG2}, \eqref{eq:NG_glob_pri} and \eqref{eq:equNG02}} with the following choice for the hyperparameters: $d_1 = d_2 = e_1 = e_2 = 0.001$, $\alpha_{a^\xi}=\alpha_{a^\tau}=5$ and $\beta_{a^\xi}=\beta_{a^\tau}=10$, \comment{implying a prior mean of  $\Ew{a^\xi}= \Ew{a^\tau}=0.1$}.
The error is assumed to be homoscedastic, with prior defined in \comment{Equation}~\eqref{eq:priorsigma} and hyperparameters $c_0 = 2.5$, $g_0 = 5$, and $G_0 = g_0/(c_0 - 1)$.

The only compulsory argument is an object of class \ ``formula'', which most users will be familiar with (see, for example, the use in the function \code{lm} in the package \pkg{stats} \citep{R}). The second argument is an optional data frame, containing the response variable and the covariates. Exemplary usage of this function is given in the code snippet below, along with the default output.
All code was on run on a personal computer with an Intel i5-8350U CPU.

\begin{CodeChunk}
\begin{CodeInput}
R> library("shrinkTVP")
R> 
R> set.seed(123)
R> sim <- simTVP(theta = c(0.2, 0, 0), beta_mean = c(1.5, -0.3, 0))
R> data <- sim$data
R> res <- shrinkTVP(y ~ x1 + x2, data = data)
\end{CodeInput}
\begin{CodeOutput}
0%   10   20   30   40   50   60   70   80   90   100%
[----|----|----|----|----|----|----|----|----|----|
**************************************************|
Timing (elapsed): 3.403 seconds.
4408 iterations per second.

Converting results to coda objects and summarizing draws... Done!
\end{CodeOutput}
\end{CodeChunk}

Note that the data in the example is generated by the function \code{simTVP}, which can create synthetic datasets of varying sizes for illustrative purposes. The inputs \code{theta} and \code{beta} can be used to specify the true $\theta_1, \ldots, \theta_d$ and $\beta_1, \ldots, \beta_d$ used in the data generating process, in order to evaluate how well \code{shrinkTVP} recaptures these true values. The values correspond to the ones used in the synthetic example of \cite{bit-fru:ach}.

The user can specify the following MCMC algorithm parameters:  \code{niter}, which determines the number of MCMC iterations including the burn-in,
\code{nburn}, which equals the number of MCMC iterations discarded as burn-in, and
\code{nthin}, indicating the thinning parameter, meaning that every nthin-th draw is kept and returned.
The default values are \code{niter = 10000}, \code{nburn = round(niter/2)} and \code{nthin = 1}.

\commentred{The user is strongly encouraged to check convergence of the  produced Markov chain, especially for a large number of covariates. The output is made \pkg{coda} compatible, so that the user can utilize the tools provided by the excellent \proglang{R} package  to assess convergence.}

\subsection{Specifying the priors}  \label{sec3:priors}

More granular control over the prior setup can be exercised by \revised{passing additional arguments to \code{shrinkTVP}. The most important argument in this regard is \code{mod\_type}, which is used to specify whether the normal-gamma-gamma (\code{mod\_type = "triple"}), the normal-gamma (\code{mod\_type = "double"}) or ridge regression (\code{mod\_type = "ridge"}) is used. Beyond this, the user can specify the hyperparameters given in Section~\ref{sec:priors} and has the possibility to fix one or both of the values of the global shrinkage \comment{parameters} ($\kappa_B^2$, $\lambda_B^2$) and the pole and tail parameters ($a^\tau$, $a^\xi$, $c^\tau$, $c^\xi$). By default, these parameters are learned from the data. The benefit of this flexibility is twofold: on the one hand, desired degrees of sparsity and global shrinkage can be achieved through fixing the hyperparameters; on the other hand, interesting special cases arise from setting certain values of hyperparameters. Under an NGG prior, for example, setting the pole and tail parameters equal to $1/2$ results in a horseshoe prior on the \comment{$\sqrt\theta_j$'s} and the \comment{$\beta_j$'s}, respectively. If the user desires a higher degree of sparsity, this can be achieved by setting the pole parameters to a value \comment{closer} to zero.  Table~\ref{tab:tablepriors} gives an overview of different model specifications. Note that different hyperparameter values can be chosen for the variances and the means of the initial values.}

\revised{
\begin{table}[]
	\setlength{\tabcolsep}{5pt}
	\centering
	\scriptsize
	\begin{tabular}{@{}lllllll@{}}
	\toprule
	& \multicolumn{3}{l}{Shrinkage on $\sqrt{\theta_j}$} & \multicolumn{3}{l}{Shrinkage on $\beta_j$} \\
	\cmidrule{2-7} 
	& $c^\xi$ & $a^\xi$ & $\kappa_B^2$ & $c^\tau$ & $a^\tau$ & $\lambda_B^2$ \\   
	\cmidrule (r{4pt}){2-4} \cmidrule (l){5-7} 
	\textit{NGG prior} \\
	Fully hierarchical NGG & $\mathcal{B}\left(\alpha_{c^\xi}, \beta_{c^\xi}\right)$ & $\mathcal{B}\left(\alpha_{a^\xi}, \beta_{a^\xi}\right)$ & $F(2a^\xi, 2c^\xi)$ & $\mathcal{B}\left(\alpha_{c^\tau}, \beta_{c^\tau}\right)$ & $\mathcal{B}\left(\alpha_{a^\tau}, \beta_{a^\tau}\right)$ & $F(2a^\tau, 2c^\tau)$ \\
	Hierarchical NGG & fixed & fixed & $F(2a^\xi, 2c^\xi)$ & fixed & fixed & $F(2a^\tau, 2c^\tau)$ \\
	NGG & fixed & fixed & fixed & fixed & fixed & fixed \\
	Hierarchical Horseshoe & fixed at 0.5 & fixed at 0.5 & $F(2a^\xi, 2c^\xi)$ & fixed at 0.5 & fixed at 0.5 & $F(2a^\tau, 2c^\tau)$ \\
  	Horseshoe & fixed at 0.5 & fixed at 0.5 & fixed & fixed at 0.5 & fixed at 0.5 & fixed \\
	\textit{NG prior} \\
	Fully hierarchical NG & \textit{-} & $\mathcal{G}(\alpha_{a^\xi}, \alpha_{a^\xi} \beta_{a^\xi})$ & $\mathcal{G}(d_1, d_2)$ & - & $\mathcal{G}(\alpha_{a^\tau}, \alpha_{a^\tau} \beta_{a^\tau})$ & $\mathcal{G}(e_1, e_2)$ \\
	Hierarchical NG & - & fixed & $\mathcal{G}(d_1, d_2)$ & - & fixed & $\mathcal{G}(e_1, e_2)$ \\
	NG & - & fixed & fixed & - & fixed & fixed\\
	Bayesian Lasso & - & fixed at 1 & fixed & - & fixed at 1 & fixed \\
	\\
	\textit{Ridge regression} & - & - & fixed & - & - & fixed \\
\bottomrule
	\end{tabular}
	\caption{Overview of different possible model specifications. Note that in the NGG prior case, the priors on the hyperparameters are scaled (e.g. $2a^\xi \sim \mathcal{B}(\alpha_{a^\xi}, \beta_{a^\xi})$). These scalings are omitted from this table for the sake of brevity. See Section~\ref{sec:priors} for details.}
	\label{tab:tablepriors}
\end{table}
}

In the following, we give some examples of models that can be estimated with the \pkg{shrinkTVP} package. In particular, we demonstrate how certain combinations of input arguments correspond to different model specifications. \revised{If the learning of a parameter is deactivated and no specific fixed value is provided, \code{shrinkTVP} will resort to default values. These equate to 0.1 for the pole and tail parameters and 20 for the global shrinkage parameters.} Note that in the following snippets of code, the argument \code{display_progress} is always set to \code{FALSE}, in order to suppress the progress bar and other outputs.

\paragraph{Fixing the \revised{pole parameters}}

It is possible \comment{to set} 
the \revised{pole parameter} $a^\xi$($a^\tau$) \comment{to a fixed value} through the input argument \code{a_xi} (\code{a_tau}), after setting \code{learn_a_xi} (\code{learn_a_tau}) to \code{FALSE}. As an example, we show how to fit a hierarchical Bayesian Lasso, both on the $\sqrt{\theta_j}$'s and on the $\beta_j$'s:

\begin{CodeInput}
R> res_hierlasso <- shrinkTVP(y ~ x1 + x2, data = data,
+    learn_a_xi = FALSE,  learn_a_tau = FALSE,
+    a_xi = 1, a_tau = 1, display_progress = FALSE)
\end{CodeInput}

\paragraph{Fixing the global shrinkage parameters}
The user can choose to fix the value of $\kappa_B^2$($\lambda_B^2$) by specifying the argument \code{kappa2_B} (\code{lambda2_B}), after setting \code{learn_kappa2_B} (\code{learn_lambda2_B}) to \code{FALSE}. In the code below, we give an example on how to fit a (non-hierarchical) Bayesian Lasso on both $\sqrt{\theta_j}$'s and $\beta_j$'s, with corresponding global shrinkage parameters fixed both to $100$:

\begin{CodeInput}
R> res_lasso <- shrinkTVP(y ~ x1 + x2, data = data,
+    learn_a_xi = FALSE, learn_a_tau = FALSE, a_xi = 1, a_tau = 1,
+    learn_kappa2_B = FALSE, learn_lambda2_B = FALSE, 
+    kappa2_B = 100, lambda2_B = 100,
+    display_progress = FALSE)
\end{CodeInput}

\revised{\paragraph{Changing the prior type} To change the model type, the input argument \code{mod\_type} has to be supplied. It has to be a string equal to either \code{"triple"}, \code{"double"} or \code{"ridge"}. As an example, we fit a hierarchical NGG prior, both on the $\sqrt{\theta_j}$'s and on the $\beta_j$'s:}
\begin{CodeInput}
R> res_tg <- shrinkTVP(y ~ x1 + x2, data = data,
+    mod_type = "triple",
+    display_progress = FALSE)
\end{CodeInput}

\revised{\paragraph{Fixing the tail parameters} Much like the pole parameters, the tail parameter $c^\xi$ ($c^\tau$) can also be fixed to a value. This is done by setting} \code{learn_c_xi} (\code{learn_c_tau}) \revised{to \code{FALSE} and then supplying the input parameter} \code{c_xi} (\code{c_tau}). \revised{As an example, the code below fits a non-hierarchical horseshoe prior, both on the $\sqrt{\theta_j}$'s and on the $\beta_j$'s:}
\begin{CodeInput}
R> res_hs <- shrinkTVP(y ~ x1 + x2, data = data,
+    mod_type = "triple",
+    learn_a_xi = FALSE, learn_a_tau = FALSE, a_xi = 0.5, a_tau = 0.5,
+    learn_c_xi = FALSE, learn_c_tau = FALSE, c_xi = 0.5, c_tau = 0.5,
+    learn_kappa2_B = FALSE, learn_lambda2_B = FALSE,
+    display_progress = FALSE)
\end{CodeInput}

\subsection{Stochastic volatility specification}

The stochastic volatility specification defined in Equation~\eqref{eq:svht} can be used by setting the option \code{sv} to \code{TRUE}. This is made possible by a call to the \code{update_sv} function exposed by the \pkg{stochvol} package.
The code below fits a model with an NG prior in which all the parameters are learned and the observation equation errors are modeled through stochastic volatility:

\begin{CodeInput}
R> res_sv <- shrinkTVP(y ~ x1 + x2, data = data, sv = TRUE,
+    display_progress = FALSE)
\end{CodeInput}

The priors on the SV parameters are the ones defined in Equation~\eqref{eq:volpriors}, with hyperparameters fixed to
$b_\mu = 0$ , $B_\mu = 1$, $\aphi = 5$, $\bphi = 1.5$ , and $\Bsv = 1$.

\subsection{Specifying the hyperparameters}

Beyond simply switching off parts of the hierarchical structure of the prior setup, users can also modify the hyperparameters governing the \revised{hyperprior} distributions. This can be done through the arguments \code{hyperprior_param} and \code{sv_param}, which both have to be named lists. \revised{Hyperparameters} not specified by the user will be set to default values, \revised{which can be found in the help file of the \code{shrinkTVP} function. Note, however, that the dependence structure (e.g. $\kappa^2_B$ depends on  $a^\xi$ and $c^\xi$ in the NGG specification) can not be changed. As such, if the user desires to change the hyperparameters of a prior that depends on other parameters, this can only be achieved by deactivating the learning of the parameters higher up in the hierarchy and fixing them to specific values. 
To demonstrate how to change specific hyperparameters, the code below modifies those governing the prior on $a^\xi$:}

\begin{CodeInput}
R> res_hyp <- shrinkTVP(y ~ x1 + x2, data = data,
+    hyperprior_param = list(beta_a_xi = 5, alpha_a_xi = 10),
+    display_progress = FALSE)
\end{CodeInput}

\revised{
	\subsection{Tuning the Metropolis-Hastings steps}
	The Metropolis-Hastings algorithm discussed in Section~\ref{sec:MCMC} can be tuned via the argument \code{MH\_tuning}. Similar to \code{hyperprior\_param} and \code{sv\_param}, it is a named list where values that are not supplied are replaced by standard values. By default, adaptive Metropolis-within-Gibbs is activated for all parameters learned from the data that requrire a Metropolis-Hastings step. Below is an example where the adaptive Metropolis is deactivated for one of the pole parameters and slightly tuned for the other:
}
\begin{CodeInput}
R> res_MH <- shrinkTVP(y ~ x1 + x2, data = data,
+    MH_tuning = list(a_xi_adaptive = FALSE,
+      a_tau_max_adapt = 0.001,
+      a_tau_batch_size = 20),
+    display_progress = FALSE)
\end{CodeInput}
\subsection{Posterior inference: Summarize and visualize the posterior distribution}

The return value of \code{shrinkTVP} is an object of type \code{shrinkTVP}, which is a named list containing a variable number of elements, depending on the prior specification. For the \revised{default NG prior}, the values are:

\begin{enumerate}
	\itemsep0em
	\item a list holding $d$ \code{mcmc.tvp} objects (one for each $\comment{\bm{\beta}_j=(\bct{j}{0}, \ldots, \bct{j}{T})}$)
containing the parameter draws in \code{beta},
	\item the parameter draws of $\bm \beta = (\beta_1, \dots, \beta_d)$ in \code{beta_mean}, 
	\item the parameter draws of $(\sqrt{\theta_1}, \dots, \sqrt{\theta_d})$ in \code{theta_sr},
	\item the parameter draws of $\tau_1^2, \ldots, \tau_d^2$ in \code{tau2},
	\item the parameter draws of $\xi_1^2,  \ldots,\xi_d^2,$ in \code{xi2},
	\item the parameter draws of $a^{\xi}$ in \code{a_xi},
	\item the parameter draws of $a^{\tau}$ in \code{a_tau},
	\item the parameter draws for $\kappa_B^2$ in \code{kappa2_B},
	\item the parameter draws for $\lambda_B^2$ in \code{lambda2_B},
	\item the parameter draws of $\sigma^2$  in \code{sigma2},
	\item the parameter draws of $C_0$ in \code{C0},
	\item MH diagnostic values in \code{MH_diag},
	\item the prior hyperparameters in \code{priorvals},
	\item the design matrix, the response and the formula in \code{model}, 
	\item summary statistics for the parameter draws in \code{summaries}
	 and
objects required for the \code{LPDS} function in  \code{internals}.
\end{enumerate}
When some parameters are fixed by the user, the corresponding output value is omitted. \revised{Additionally, increasing or decreasing the amount of levels in the hierarchy of the prior also changes which values are returned. For example, if }\code{mod_type} \revised{is changed to} \code{"triple"} \revised{and the learning of the tail parameters $c^\xi$ and $c^\tau$ is not deactivated, then the output will also contain the respective parameter draws in }\code{c_xi} \revised{and} \code{c_tau}.
In the SV case, the draws for the parameters of the SV model on the errors are contained in \code{sv_mu}, \code{sv_phi} and \code{sv_sigma}.  For details, see \cite{kas:dea}.

The two main tools for summarizing the output of \code{shrinkTVP} are the \code{summary} and \code{plot} methods implemented for \code{shrinkTVP} objects. \code{summary} has two arguments beyond the \code{shrinkTVP} object itself, namely \code{digits} and \code{showprior}, which control the output displayed. \code{digits} indicates the number of decimal places to round the posterior summary statistics to, while \code{showprior} determines whether or not to show the prior distributions resulting from the user input. In the example below, the default \code{digits} value of 3 is used, while the prior specification is omitted. The output of \code{summary} consists of the mean, standard deviation, median, 95\% \comment{highest} posterior density region and effective sample size (ESS) for the non time-varying parameters.

\begin{CodeChunk}
\begin{CodeInput}
R> summary(res, showprior = FALSE)
\end{CodeInput}
\begin{CodeOutput}
Summary of 5000 MCMC draws after burn-in of 5000.
Statistics of posterior draws of parameters (thinning = 1):

 param                   mean   sd      median HPD 2.5% HPD 97.5% ESS 
 beta_mean_Intercept     0.164  0.432   0      -0.376   1.362     426 
 beta_mean_x1            -0.248 0.157   -0.274 -0.483   0.013     110 
 beta_mean_x2            -0.002 0.037   0      -0.105   0.069     3741
																	  
 abs(theta_sr_Intercept) 0.423  0.064   0.418  0.307    0.551     345 
 abs(theta_sr_x1)        0.013  0.024   0      0        0.067     133 
 abs(theta_sr_x2)        0.002  0.007   0      0        0.013     598 
																	  
 tau2_Intercept          6.989  151.023 0.001  0        4.85      4799
 tau2_x1                 6.615  240.363 0.096  0        4.658     5000
 tau2_x2                 0.438  15.431  0      0        0.105     5000
																	  
 xi2_Intercept           23.525 563.676 0.265  0.007    10.827    5000
 xi2_x1                  1.883  89.188  0      0        0.07      5000
 xi2_x2                  0.011  0.331   0      0        0.003     5000
																	  
 a_xi                    0.08   0.039   0.074  0.014    0.154     169 
																	  
 a_tau                   0.091  0.041   0.085  0.025    0.176     452 
																	  
 kappa2_B                29.519 104.414 1.507  0        148.318   4205
																	  
 lambda2_B               53.677 184.784 2.182  0        257.231   1102
																	  
 sigma2                  0.993  0.125   0.984  0.767    1.248     1622
																	  
 C0                      1.72   0.634   1.648  0.606    2.962     5000
\end{CodeOutput}
\end{CodeChunk}

\begin{figure}[h!]
	\centering
	\includegraphics[width=\linewidth]{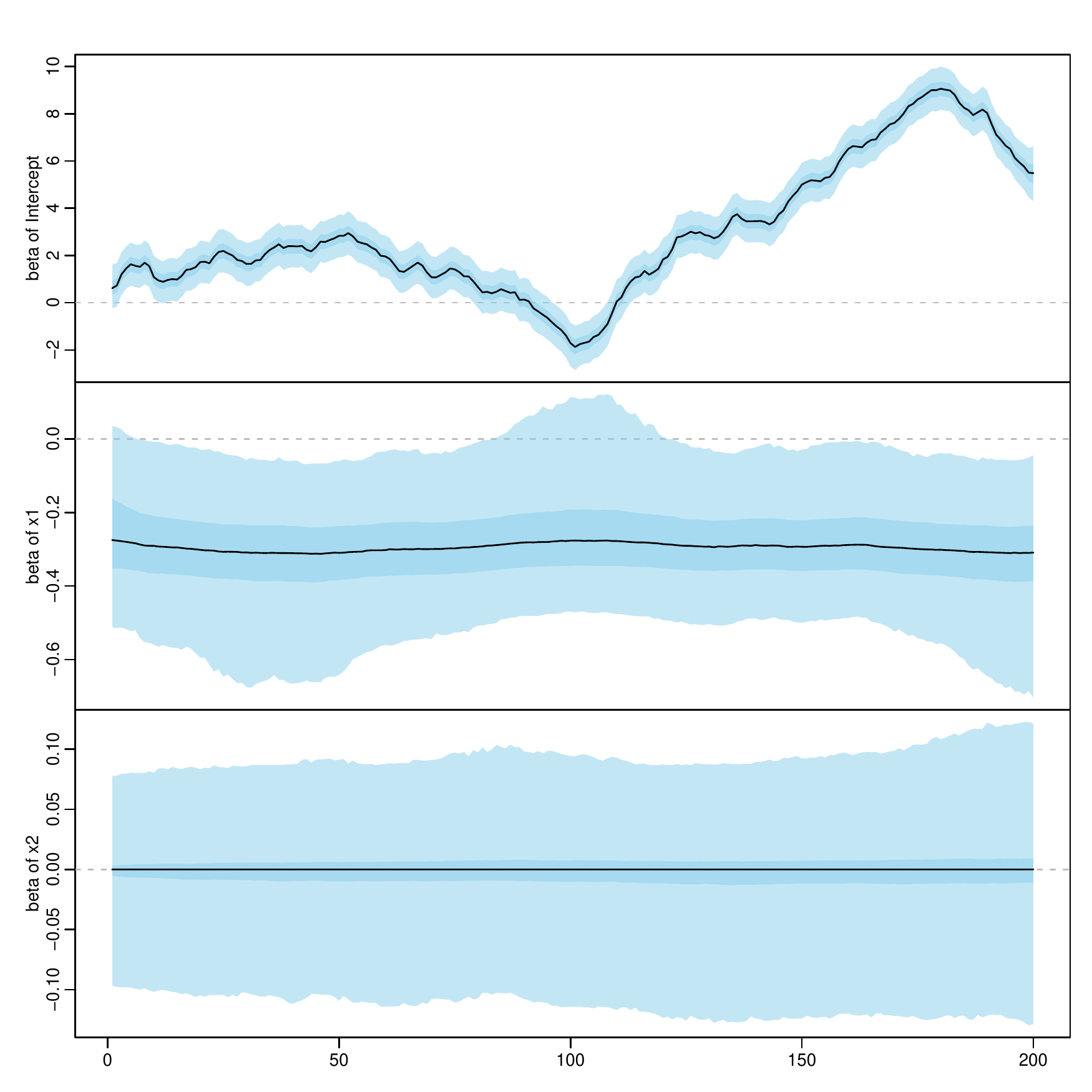}
	\caption{Visualization of the  evolution of the \comment{time-varying  parameter  $\bm{\beta}_j=(\bct{j}{0}, \ldots, \bct{j}{T}), j=1, 2, 3,$
	over time  $t=0,\ldots,T$}, as provided by the \code{plot} method. \code{plot} is in turn calling \code{plot.mcmc.tvp} on the individual \code{mcmc.tvp} objects. The median is displayed as a black line, and the shaded areas indicate the \comment{pointwise} 95\% and 50\% posterior credible intervals.}
\label{fig:beta}
\end{figure}

The \code{plot} method can be used to visualize the posterior distribution estimated by \code{shrinkTVP}. Aside from a \code{shrinkTVP} object, its main argument is \code{pars}, a character vector containing the names of the parameters to visualize. \code{plot} will call either \code{plot.mcmc.tvp} from the \pkg{shrinkTVP} package if the parameter is time-varying or \code{plot.mcmc} from the \pkg{coda} package, if the parameter is non time-varying.  The default value of \code{pars} is \code{c("beta")}, leading to \code{plot.mcmc.tvp} being called on each of the $\beta_{jt}$, for $j =1, \ldots, d$. See the code below for an example and Figure~\ref{fig:beta} for the corresponding output.

\begin{CodeInput}
R> plot(res)
\end{CodeInput}

\begin{figure}[h!]
\centering
\includegraphics[width=\linewidth]{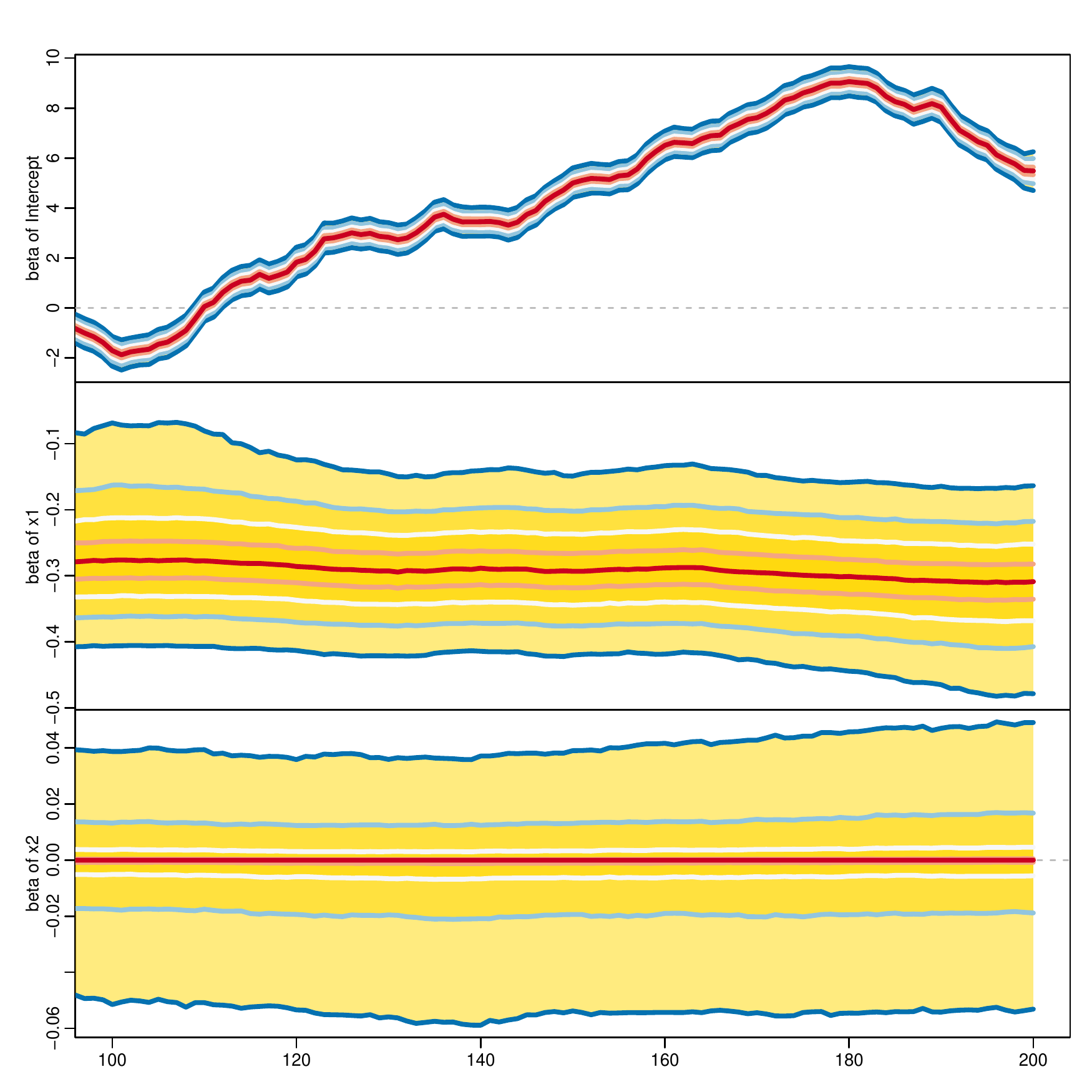}
\caption{Visualization of the  evolution of  the \comment{time-varying  parameter  $\bm \beta_{t}$ over time  $t=100,\ldots,200$ for $j=1, \ldots, 3$}. In this example, the x-axis of the plot was restricted with \code{xlim}, the color of the shaded areas was changed to yellow and colored solid lines have been added to delimit the credible intervals. The colored lines represent the median and the \comment{pointwise}	10\%, 20\%, 30\% 40\%, 60\%, 70\%, 80\%, and 90\% quantiles.}
\label{fig:beta2}
\end{figure}

The \code{plot.mcmc.tvp} method displays empirical posterior credible intervals of a time-varying parameter over time, i.e., $\beta_{jt}$, for $j =1, \ldots, d$ and $\sigma^2_t$ in the case of stochastic volatility. By default, the \comment{pointwise} 95\% and 50\% posterior credible intervals are displayed as shaded areas layered on top of one another, with the median represented by a black line, with an additional grey, dashed line at zero. To ensure that users have flexiblity in the plots created, a host of options are implemented for customisation. The bounds of the credible intervals can be modified through the \code{probs} input, allowing for different levels of uncertainty visualization. The arguments \code{quantlines} and \code{shaded} take boolean vectors as inputs, and determine if the corresponding credible intervals will be displayed through shading and/or lines. The shaded areas can be customised via the arguments \code{shadecol} and \code{shadealpha}, which determine the color and the degree of transparency of the shaded areas. The lines representing the quantiles can be adjusted through \code{quantlty}, \code{quantcol} and \code{quantlwd}, which modify the line type, color and line width, respectively. In the spirit of \proglang{R}, all of these arguments are vectorised and the supplied vectors are recycled in the typical \proglang{R} fashion if necessary. The first element of these vectors is always applied to the outermost credible interval, the second to the second outermost and so forth. The horizontal line at zero can be similarly adjusted through \code{zerolty}, \code{zerolwd} and \code{zerocol} or entirely turned off by setting \code{drawzero} equal to \code{FALSE}. All further arguments are passed on to the standard \code{plot} method, allowing for changes to the line representing the median and other plot modifications that users of \proglang{R} are familiar with. An example of possible customisation can be seen in the code below, with the corresponding output being Figure~\ref{fig:beta2}.

\begin{CodeInput}
R> library("RColorBrewer")
R> color <- brewer.pal(5, "RdBu")
R> plot(res, pars = "beta", xlim = c(100, 200),
+    probs = seq(0.1, 0.9, by = 0.1),
+    quantlines = T, quantcol = color[5:2], quantlty = 1,
+    quantlwd = 3, col = color[1], lwd = 3, shadecol = "gold1")
\end{CodeInput}

\begin{figure}[h!]
\centering
\includegraphics[width=\textwidth,height=\textheight,keepaspectratio]{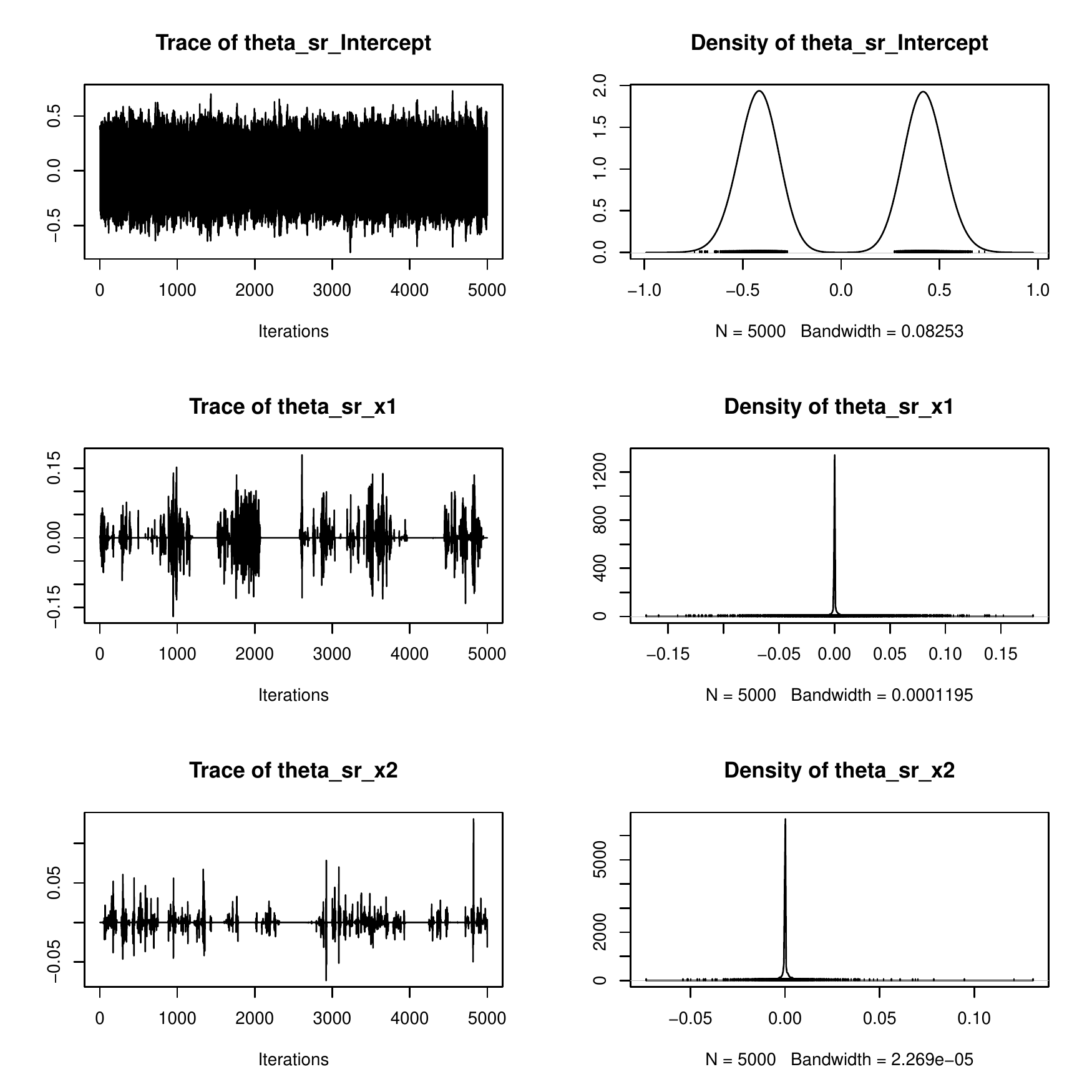}
\caption{\comment{Trace plots (left column) and kernel density estimates of the posterior density (right column) for the parameters
$\sqrt{\theta}_1, \dots, \sqrt{\theta}_3$}, as provided by the \code{plot} method. \code{plot} is in turn calling \pkg{coda}'s \code{plot.mcmc}.}
\label{fig:theta}
\end{figure}

To visualize other parameters via the \code{plot} method, the user has to change the \code{pars} argument. \code{pars} can either be set to a single character object or to a vector of characters containing the names of the parameter draws to display. In the latter case, the \code{plot} method will display groups of plots at a time, prompting the user to move on \comment{to} the next series of plots, similarly to how \pkg{coda} handles long plot outputs. Naturally, as all parameter draws are converted to \pkg{coda} objects, any
\comment{method  from this package  that users are familiar with} 
(e.g., to check convergence) can be applied to the parameter draws contained in a \code{shrinkTVP} object. An example of this can be seen in Figure~\ref{fig:theta}, where \code{pars = "theta_sr"},  \comment{changes} 
the output to a graphical summary of the parameter draws of $\sqrt{\theta_1}, \dots, \sqrt{\theta_d}$, using \pkg{coda}'s \code{plot.mcmc} function. To obtain Figure~\ref{fig:theta}, one can run

\begin{CodeInput}
R> plot(res, pars = "theta_sr")
\end{CodeInput}

\section{Predictive performances and model comparison}
\label{sec:LPDS}

Within a Bayesian framework, a natural way to predict a future observation is through its posterior predictive density.  For this reason, log-predictive density scores (LPDSs) provide a means of assessing how well the model performs in terms of prediction on real data. The log-predictive density score for time $t_0 +1$
is obtained by evaluating at $y_{t_0 +1}$ the
\comment{log of the posterior predictive density} 
obtained by fitting the model to the previous $t_0$ data points.
Given the data up to time $t_0$, the posterior predictive density at time $t_0 + 1$ is given by
\begin{align} \label{MCpred}
p(y_{t_0 + 1}| y_{1}, \ldots, y_{t_0}, \bm x_{t_0 +1} ) =  \int p(y_{t_0 + 1}| \bm x_{t_0 +1}, \bm \psi )  p (\bm \psi| y_{1}, \ldots, y_{t_0} )d \bm \psi,
\end{align}
where $\bm \psi$ is the set of model parameters \comment{and latent variables up to $t_0+1$.  For a TVP model with homoscedastic  errors, $\bm \psi=  (\tilde {\bm \beta}_0,\ldots \tilde {\bm \beta}_{t_0 +1}, \sqrt \theta_1, \ldots, \sqrt{\theta_d}, \beta_1, \ldots, \beta_d, \sigma^2)$,
whereas for a TVP model with  SV errors,
 $\bm \psi=  (\tilde {\bm \beta}_0,\ldots \tilde {\bm \beta}_{t_0 +1}, \sqrt \theta_1, \ldots, \sqrt{\theta_d}, \beta_1, \ldots, \beta_d,
 \sigma^2_1, \ldots,\sigma^2_{t_0 +1})$.}
\comment{Given $M$ samples from the posterior distribution of the parameters and latent variables, $p (\bm \psi| y_{1}, \ldots, y_{t_0} )$, Monte Carlo integration
could be applied immediately to approximate (\ref{MCpred}).
However, \cite{bit-fru:ach} propose a more efficient approximation of the predictive density,   \comment{the so-called conditionally optimal Kalman mixture approximation which is}
obtained by analytically integrating out $\tilde {\bm \beta}_{t_0+1}$ from the likelihood at time $t_0 +1$.}

 In the homoscedastic error case, given $M$ samples from the posterior distribution of the parameters  \comment{and the latent variables up to $t_0$},
 \commentred{Monte Carlo integration  of  the resulting predictive density yields following   mixture approximation,
\begin{eqnarray} \label{eq:approx_mixture}
\hspace{-0.15cm}
 p(y_{t_0+1}| y_{1},\ldots, y_{t_0}, \xv_{t_0+1}) &\approx& \dfrac{1}{M} \sum_{m=1}^M
\Normalpdfa{y_{t_0+1}}{\yhat _{t_0+1} \imm{m}, \yS _{t_0+1} \imm{m} }, \\
 \yhat \imm{m} _{t_0+1}& =& \bm x_{t_0+1} \bm \beta^{(m)} + \bm F_{t_0+1}\imm{m}  \bm m_{t_0} \imm{m}, \nonumber  \\
 \yS \imm{m} _{t_0+1} &= & \bm F_{t_0+1} \imm{m} (\SSigma_{t_0} \imm{m} + I_d) ( \bm F_{t_0+1}\imm{m}) ^\top + (\sigma^{2})\imm{m} , \nonumber
\end{eqnarray}
 where  the conditional predictive  densities  are  Gaussian and   the  conditional  moments   depend on the MCMC draws.
 The   mean  $\yhat _{t_0+1} \imm{m}$  and  the variance  $\yS _ {t_0+1} \imm{m} $}
are computed for  the  $m$th MCMC iteration from
 $\bm F_{t_0+1}=  \bm x_{t_0+1} \text{Diag}(\sqrt{\theta_1}, \ldots, \sqrt{\theta_d)}$ and
 the mean  $\bm m_{t_0}$ and  the covariance matrix  $\SSigma_{t_0}$  of the posterior distribution of $\tilde {\bm \beta}_{t_0}$.
These quantities can be obtained by iteratively calculating $\SSigma_{t}$ and $\bm m_{t}$ up to time $t_0$, as described in \cite{mcc-etal:sim}:
\begin{align*}
&\SSigma_1 = (\OOmega_{11})^{-1}, \qquad \bm m_1 = \SSigma_1 \bm c_1,\\
&\SSigma_t = (\OOmega_{tt} - \OOmega_{t-1,t}^{\top} \SSigma_{t-1} \OOmega_{t-1,t})^{-1}, \qquad \bm m_t = \SSigma_t (\bm c_t - \OOmega_{t-1,t}^{\top} \bm m_{t-1} ).
\end{align*}
The quantities $\cv_t$,  $\OOmega_{tt}$ and  $\OOmega_{t-1,t}$ for $t=1,  \ldots, t_0$ are given in Appendix~\ref{sec:mccau}.

For the SV case,  it is still  possible to analytically integrate  out $\tilde {\bm \beta}_{t_0+1}$ from the likelihood at time $t_0 +1$ conditional on
a known value of $\sigma^2_{t_0+1}$, however it is not possible to integrate  the likelihood with respect to both  latent variables $\tilde {\bm \beta}_{t_0+1}$  and $\sigma^2_{t_0+1}$.
Hence, at each MCMC iteration a draw is taken from the predictive distribution of  $\sigma^2_{t_0+1}=\exp (h_{t_0+1})$, derived from Equation~\eqref{eq:svht},  and used to calculate the \comment{conditional predictive density of   $y_{t_0+1}$}.
The approximation of the one-step ahead predictive density can then be obtained through the following steps:
\begin{enumerate}
	\item    for each MCMC draw of $(\mu, \phi,\sigma_{\eta}^2) \imm{m}$ and $h_{t_0} \imm{m}$,  obtain a  draw of $(\sigma^{2}_{t_0+1})\imm{m}$; 	\item calculate the   conditionally optimal Kalman   mixture approximation as \commentred{in (\ref{eq:approx_mixture}) with following
slightly different values $\yS _ {t_0+1} \imm{m} $:}
\commentred{	\begin{align*}
&  \yS _ {t_0+1} \imm{m} =  \bm F_{t_0+1}  \imm{m} (\SSigma_{t_0} \imm{m} + I_d) ( \bm F_{t_0+1} \imm{m}) ^\top +  (\sigma^{2}_{t_0+1})\imm{m}, 		\end{align*}
	where $\bm F_{t_0+1}$ and $\SSigma_{t_0}$ are the same as defined above.}
\end{enumerate}

These calculations can be performed by the \code{LPDS} function, based on a fitted TVP model resulting from a call to \code{shrinkTVP}. The function's arguments are an object of class \code{shrinkTVP} and  \code{data_test},  a data frame with one row, containing covariates and response at time $t_0 + 1$. The following snippet of code fits a \code{shrinkTVP} model to synthetic data up to $T - 1$, and then calculates the LPDS at time $T$. The obtained LPDS score is then displayed.
For an example on how to calculate LPDSs for $k$ points in time, please see~Section~\ref{sec:usmacro}. 

\begin{figure}[t!]
	\centering
	\includegraphics[width=0.6\linewidth]{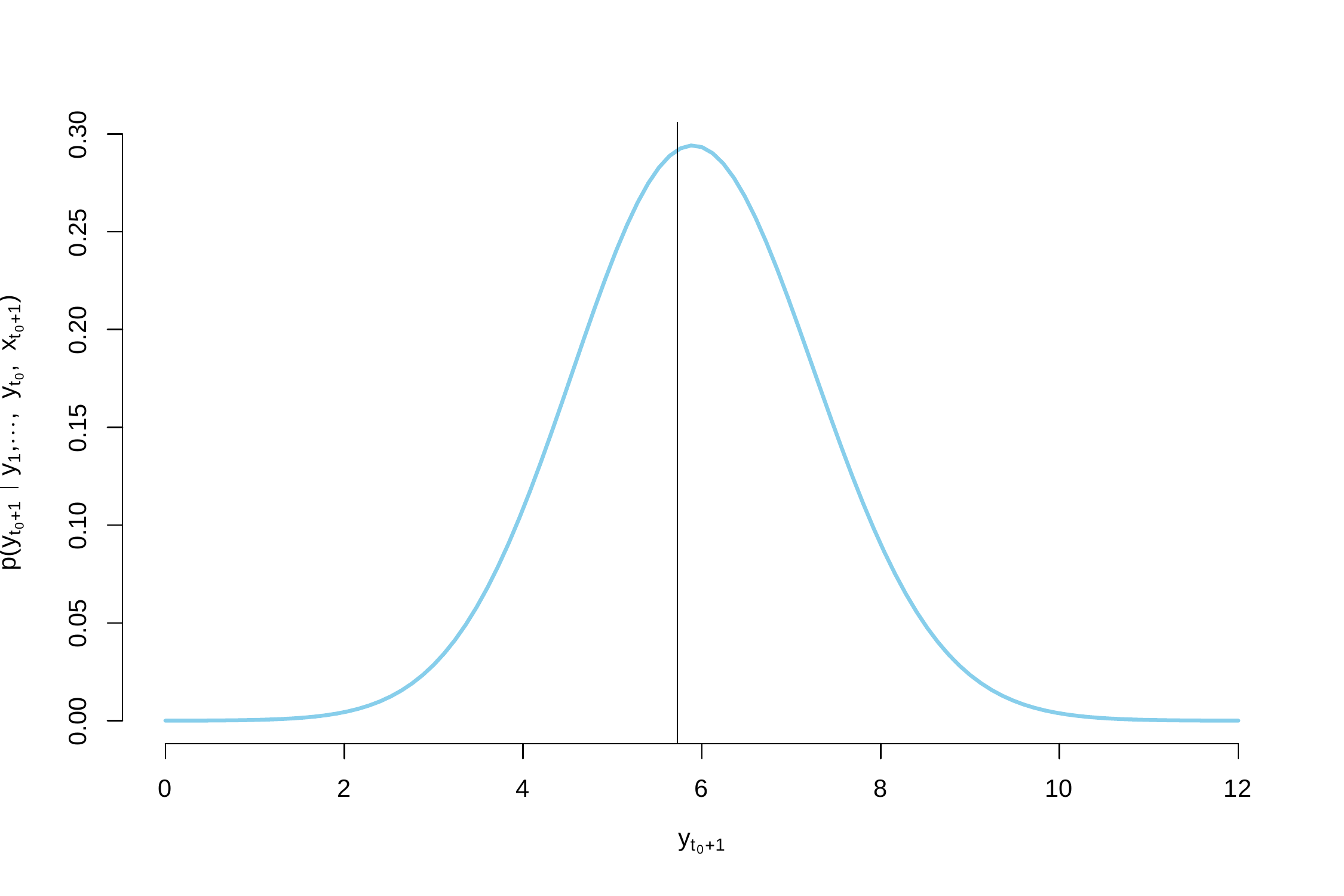}
	\caption{{One-step ahead predictive density $p(y_{t_0+1}| y_{1},\ldots, y_{t_0}, \xv_{t_0+1}) $ for a synthetic data set. The black vertical line represents the true realisation of $y_{{t_0}+1}$}.}
	\label{fig:pred_dens}
\end{figure}

\begin{CodeChunk}
\begin{CodeInput}
R> res_LPDS <- shrinkTVP(y ~ x1 + x2, data = data[1:(nrow(data) - 1),],
+    display_progress = FALSE)
R> LPDS(res_LPDS, data[nrow(data), ])
\end{CodeInput}
\begin{CodeOutput}
[1] -1.231744
\end{CodeOutput}
\end{CodeChunk}

An additional functionality provided by the package \pkg{shrinkTVP} is the evaluation of the one-step ahead predictive density through the function

\code{eval_pred_dens}.

It takes as inputs an object of class \code{shrinkTVP}, a one row data frame containing $\bm x_{t_0+1}$ and a point, or vector of points, at which the predictive density is to be evaluated. It returns a vector of the same length, containing the value of the density at the points the user supplied. An example of this can be seen in the code below. 

\begin{CodeChunk}
\begin{CodeInput}
R> eval_pred_dens(1:3, res_LPDS, data[nrow(data), ])
\end{CodeInput}
\begin{CodeOutput}
[1] 0.0004023221 0.0043769491 0.0285444188
\end{CodeOutput}
\end{CodeChunk}

Thanks to its vectorised nature, \code{eval_pred_dens}  
can be plugged directly into functions that expect an expression that evaluates to the length of the input, such as the \code{curve} function from the \pkg{graphics} \citep{R} package. The following snippet of code exploits this behaviour to plot the posterior predictive density. The result can be seen in Figure~\ref{fig:pred_dens}.

\begin{CodeChunk}
\begin{CodeInput}
R> curve(eval_pred_dens(x, res_LPDS, data[nrow(data), ]), to = 12,
+    ylab = bquote("p(" * y[t[0]+1] * "\uff5c" * y[1] * ","
+    * ldots * "," ~ y[t[0]] * "," ~ x[t[0]+1] * ")"),
+    xlab = expression(y[t[0]+1]), lwd = 2.5, col = "skyblue", axes = FALSE)
R> abline(v = data$y[nrow(data)])
R> axis(1)
R> axis(2)
\end{CodeInput}
\end{CodeChunk}

\section{Predictive exercise: usmacro dataset}
\label{sec:usmacro}

In the following, we provide a brief demonstration on how to use the \pkg{shrinkTVP} package on real data and compare different prior specifications via LPDSs. Specifically, we consider the \code{usmacro.update} dataset from the \pkg{bvarsv} package \citep{kru:bva}. The dataset \code{usmacro.update} contains the inflation rate, unemployment rate and treasury bill interest rate for the United States, from 1953:Q1 to 2015:Q2, \comment{that is $T=250$}. The same dataset up to 2001:Q3 was used by \cite{pri:tim}. The response variable is the inflation rate \code{inf}, while the predictors are the lagged inflation rate \code{inf_lag}, the lagged unemployed rate \code{une_lag} and the lagged treasury bill interest \code{tbi_lag}. We construct our dataset as follows:

\begin{CodeInput}
R> library("bvarsv")
R> data("usmacro.update")
R>
R> # Create matrix of lags and create final data set
R> lags <- usmacro.update[1:(nrow(usmacro.update) - 1), ]
R> colnames(lags) <- paste0(colnames(lags), "_lag")
R> us_data <- data.frame(inf = usmacro.update[2:nrow(usmacro.update), "inf"],
+    lags)
\end{CodeInput}

In the snippet of code below, we \revised{estimate a TVP model with a fully hierarchical NG prior}
for $60000$ iterations, with a thinning of $10$ and a burn-in of $10000$, hence keeping $5000$ posterior draws.

\begin{CodeInput}
R> us_res <- shrinkTVP(inf ~ inf_lag + une_lag + tbi_lag, us_data,
+    niter = 60000, nburn = 10000, nthin = 10,
+    display_progress = FALSE)
\end{CodeInput}

\begin{figure}[t!]
	\centering
	\includegraphics[width=\linewidth]{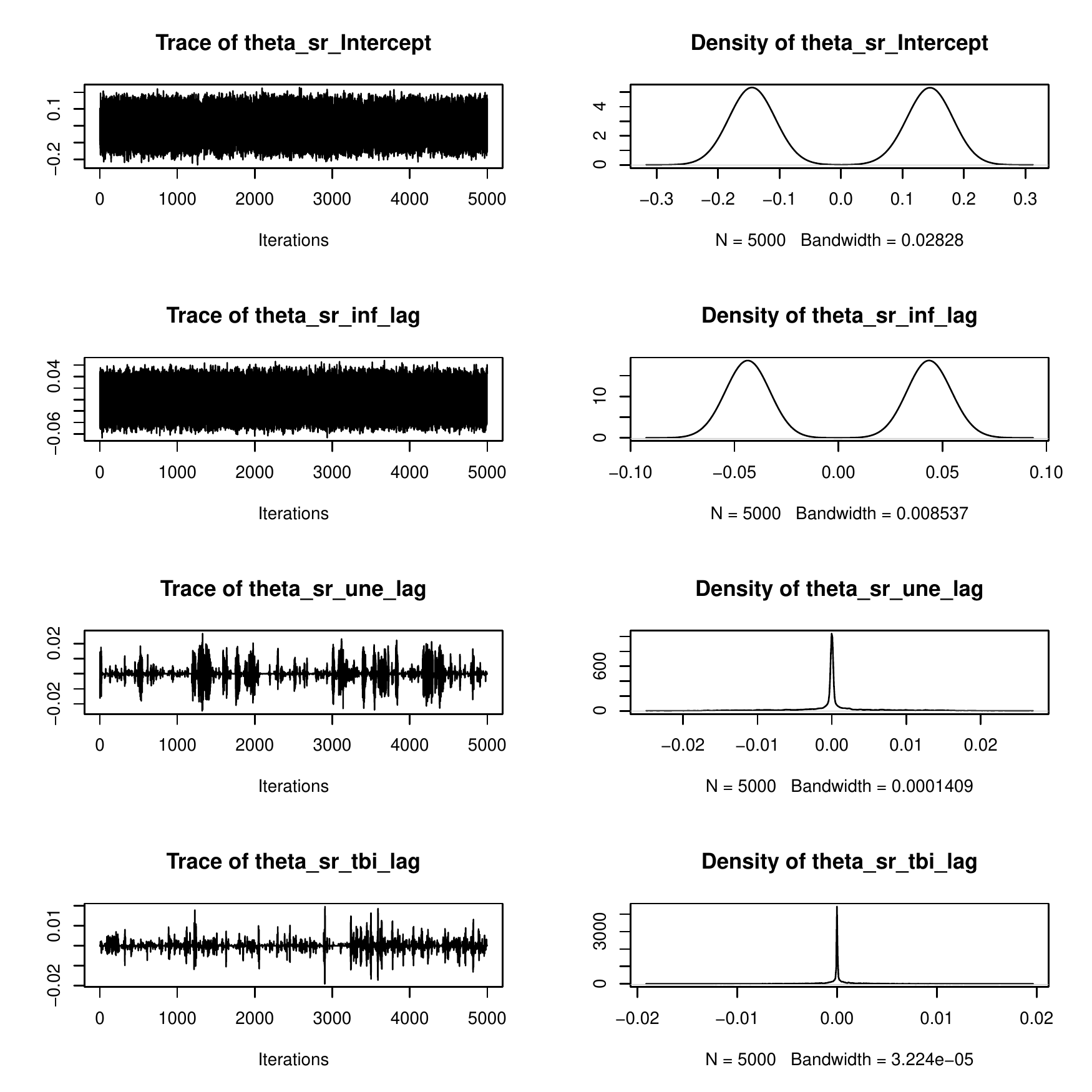}
	\caption{\comment{Trace plots (left column) and kernel density estimates of the posterior density (right column) for the parameters} $\sqrt{\theta_1}, \ldots, \sqrt{\theta_4}$ for the \code{usmacro.update} dataset.}
	\label{fig:theta_us}
\end{figure}

Once we have fit the model, we can perform posterior inference by using the \code{summary} and \code{plot} methods. The summary is shown below, while Figure~\ref{fig:beta_us} shows the paths of $\bm \beta_t$ evolving over time, and Figure~\ref{fig:theta_us} displays the trace plots (left column) and posterior densities (right column) of $\sqrt{\theta_1}, \ldots, \sqrt{\theta_4}$ obtained via the \code{plot} method.

\begin{CodeChunk}
\begin{CodeInput}
R> summary(us_res, showprior = FALSE)
\end{CodeInput}
\begin{CodeOutput}
Summary of 50000 MCMC draws after burn-in of 10000.
Statistics of posterior draws of parameters (thinning = 10):

 param                   mean    sd        median HPD 2.5% HPD 97.5% ESS 
 beta_mean_Intercept     0.415   0.436     0.326  -0.132   1.266     639 
 beta_mean_inf_lag       0.733   0.191     0.742  0.347    1.093     756 
 beta_mean_une_lag       -0.144  0.059     -0.149 -0.234   0.002     345 
 beta_mean_tbi_lag       0.008   0.022     0      -0.02    0.065     661 
																		 
 abs(theta_sr_Intercept) 0.144   0.025     0.145  0.098    0.195     1003
 abs(theta_sr_inf_lag)   0.044   0.006     0.044  0.031    0.056     2525
 abs(theta_sr_une_lag)   0.003   0.005     0      0        0.014     117 
 abs(theta_sr_tbi_lag)   0.001   0.002     0      0        0.006     478 
																		 
 tau2_Intercept          354.575 24026.311 0.158  0        15.251    5000
 tau2_inf_lag            325.535 15149.831 0.793  0        36.284    5000
 tau2_une_lag            20.052  729.984   0.049  0        3.887     5000
 tau2_tbi_lag            30.132  1447.597  0      0        0.057     5000
																		 
 xi2_Intercept           1.546   30.218    0.032  0        0.986     5000
 xi2_inf_lag             0.292   5.131     0.005  0        0.268     5000
 xi2_une_lag             0.014   0.717     0      0        0.005     5000
 xi2_tbi_lag             0.002   0.048     0      0        0.001     5000
																		 
 a_xi                    0.095   0.039     0.09   0.029    0.171     1242
																		 
 a_tau                   0.087   0.042     0.082  0.015    0.167     2290
																		 
 kappa2_B                120.337 250.836   22.127 0        609.238   5000
																		 
 lambda2_B               7.123   22.902    0.601  0        33.607    3324
																		 
 sigma2                  0.018   0.006     0.017  0.007    0.029     1140
																		 
 C0                      0.126   0.062     0.114  0.027    0.247     2360
\end{CodeOutput}
\end{CodeChunk}

\begin{figure}[t!]
	\centering
	\includegraphics[width=\linewidth]{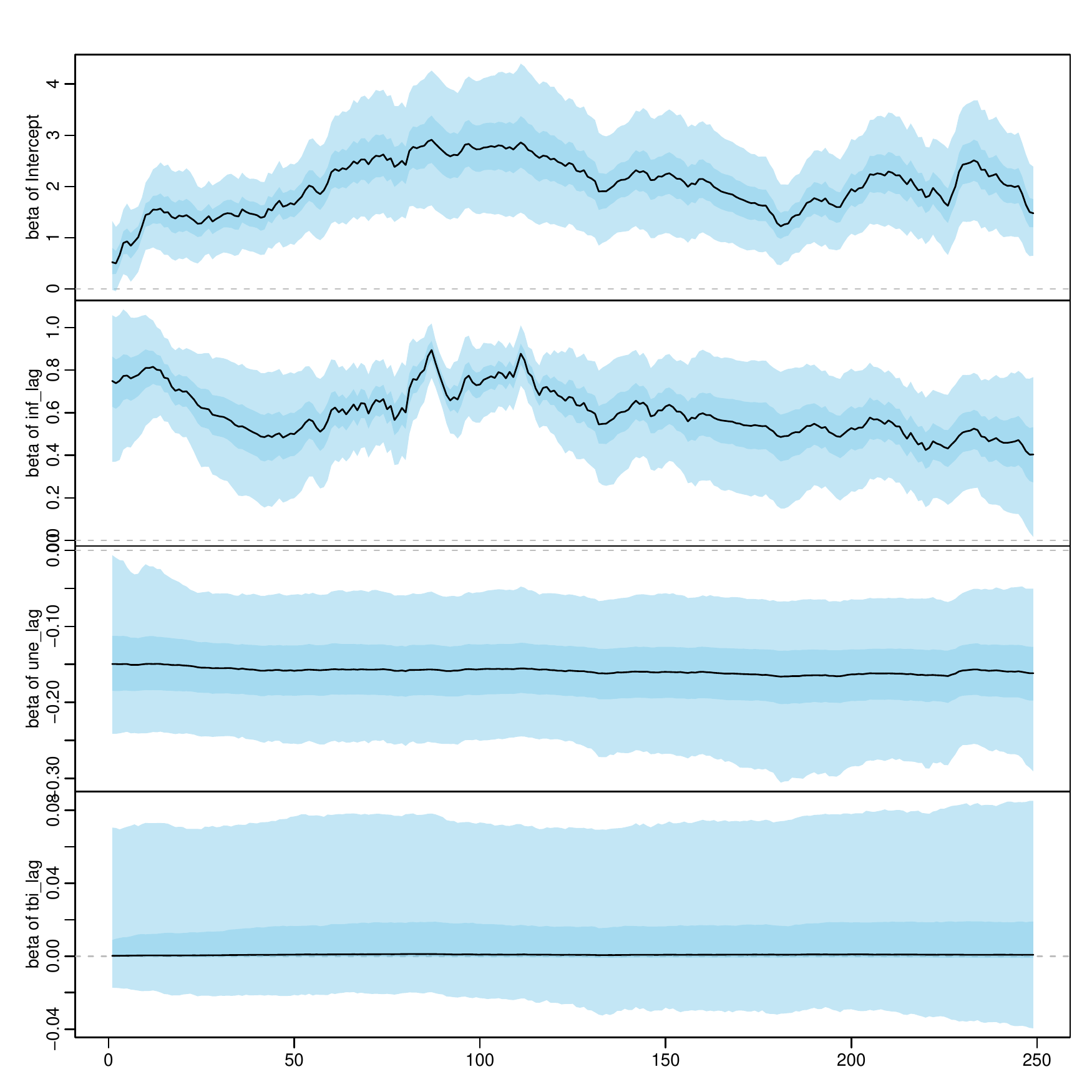}
	\caption{Visualization of the evolution of  \comment{the time-varying  parameter  $\bm{\beta}_j=(\bct{j}{0}, \ldots, \bct{j}{T})$ over time  $t=0,\ldots,T$  for $j=1, \ldots, 4$} 
for the \code{usmacro.update} dataset. The median is displayed as a black line, and the shaded areas indicate the \comment{pointwise} 95\% and 50\% posterior credible intervals.}
	\label{fig:beta_us}
\end{figure}

It appears clear by looking at Figure~\ref{fig:beta_us} that the intercept and the parameter associated with the lagged inflation rate are time-varying, while  the parameters associated with the lagged treasury bill interest rate and the lagged unemployment rate are relatively constant. This can be confirmed by looking at the posterior distributions of the corresponding standard deviations, displayed in Figure~\ref{fig:theta_us}. The posterior densities of the standard deviations associated with the intercept and the lagged inflation are bimodal, with very little mass around zero. This bimodality results from the non-identifiability of the sign of the standard deviation. As a convenient side effect, noticeable bimodality in the density plots of the posterior distribution
  \comment{$p(\sqrt{\theta}_j|\ym)$ of the standard deviations  $\sqrt{\theta}_j$}   
  is a strong indication of time variability in the associated parameter  \comment{$\bct{j}{t}$.} 
  Conversely, the posterior densities of the standard deviations associated with the lagged unemployment and the lagged treasury bill interest rate have a pronounced spike at zero, indicating strong model evidence in favor of constant parameters. Moreover, the path of the parameter of the treasury bill interest rate is centered at zero, indicating that this parameter is neither time-varying nor significant.

\begin{figure}[h!]
	\centering
	\includegraphics[width=\linewidth]{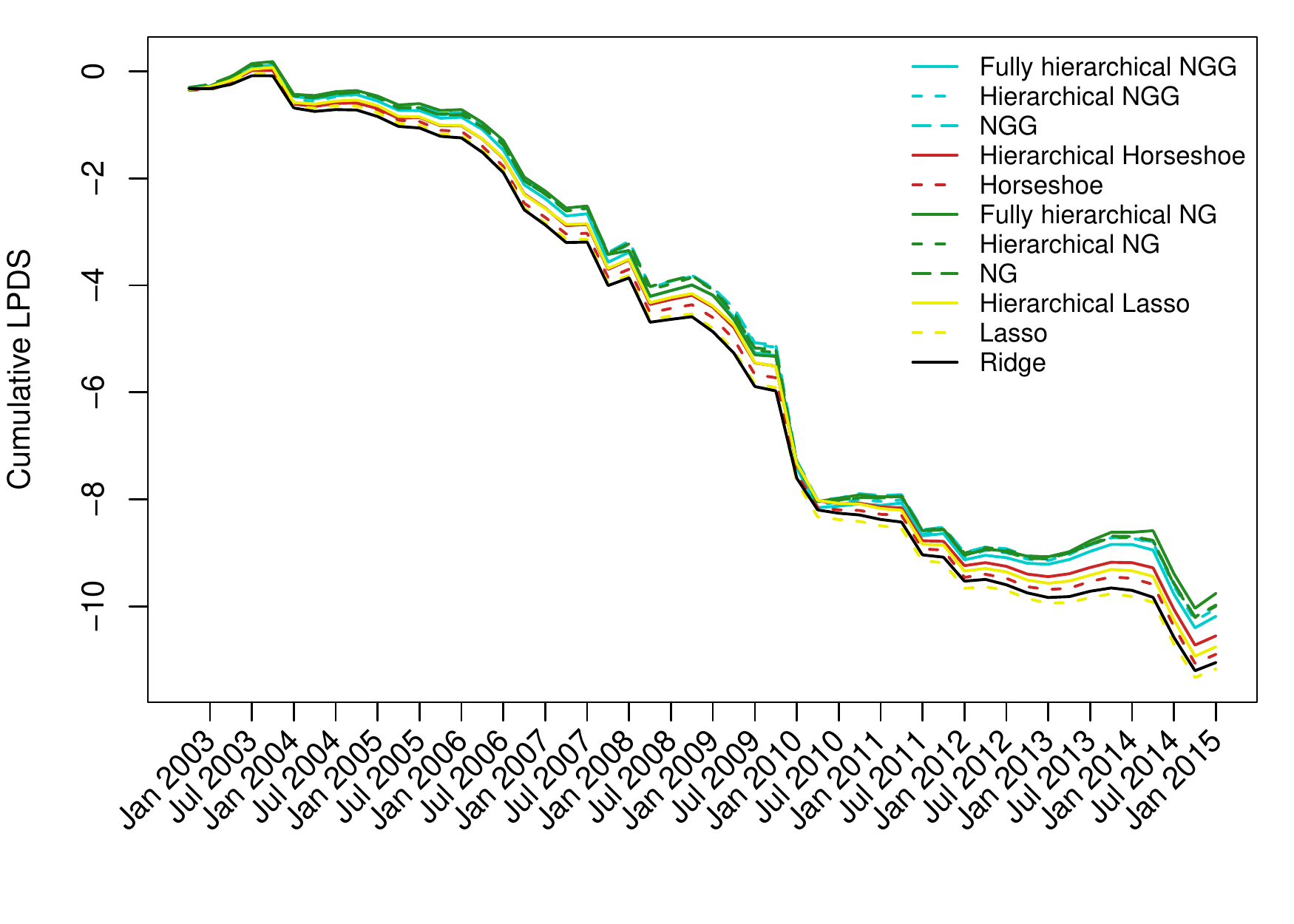}
	\captionof{figure}{Cumulative LPDSs for the last 50 quarters of the \code{usmacro.update} dataset, for \revised{eleven} different shrinkage priors: \revised{(1) the full hierarchical NGG prior, (2) the hierarchical NGG \comment{prior}  with fixed $a^\xi = a^\tau = c^\xi = c^\tau = 0.1$, (3) the NGG \comment{prior} with $a^\xi = a^\tau = c^\xi = c^\tau = 0.1$ and $\kappa_B^2 = \lambda_B^2 = 20$, (4) the hierarchical horseshoe prior, (5) the horseshoe prior $\kappa_B^2 = \lambda_B^2 = 20$, (6) the full hierarchical NG prior, (7) the hierarchical NG prior with fixed $a^\xi = a^\tau = 0.1$, (8) the NG prior with $a^\xi = a^\tau = 0.1$ and $\kappa_B^2 = \lambda_B^2 = 20$, (9) the hierarchical Bayesian Lasso, and (10) the Bayesian Lasso with $\kappa_B^2 = \lambda_B^2 = 20$ and (11) ridge regression with $\kappa_B^2 = \lambda_B^2 = 20$.}}
	\label{fig:LPDS}
\end{figure}

In order to compare the predictive performances of different shrinkage priors, we calculate one-step ahead LPDSs for the last 50  points in time for \revised{eleven} different prior choices: \revised{(1) the full hierarchical NGG prior, (2) the hierarchical NGG \comment{prior}  with fixed $a^\xi = a^\tau = c^\xi = c^\tau = 0.1$, (3) the NGG \comment{prior} with $a^\xi = a^\tau = c^\xi = c^\tau = 0.1$ and $\kappa_B^2 = \lambda_B^2 = 20$, (4) the hierarchical horseshoe prior, (5) the horseshoe prior $\kappa_B^2 = \lambda_B^2 = 20$, (6) the full hierarchical NG prior, (7) the hierarchical NG prior with fixed $a^\xi = a^\tau = 0.1$, (8) the NG prior with $a^\xi = a^\tau = 0.1$ and $\kappa_B^2 = \lambda_B^2 = 20$, (9) the hierarchical Bayesian Lasso, and (10) the Bayesian Lasso with $\kappa_B^2 = \lambda_B^2 = 20$ and (11) ridge regression with $\kappa_B^2 = \lambda_B^2 = 20$.} Figure~\ref{fig:LPDS} shows the cumulative LPDSs for the last 50 quarters of the \code{usmacro.update} dataset. \revised{The default prior, the fully hierarchical NG prior on both the $\beta_j$'s and the \comment{$\sqrt{\theta_j}$'s, performs the best in terms of prediction.}} In Appendix~\ref{sec:multicore} we show how to obtain LPDSs for different models and points in time, using the packages \pkg{foreach} \citep{wes:for} and \pkg{doParallel} \citep{wes:doP}.

\section{Conclusions}
\label{sec:conclusions}

The goal of this  paper was to introduce the reader to the functionality of the \proglang{R} package \pkg{shrinkTVP} \citep{kna-etal:shr}. This \proglang{R} package provides a fully Bayesian approach for statistical inference in TVP models with shrinkage priors.
\comment{On the one hand, the package provides an easy entry point for users
who want  to pass on only their data in a first step of exploring  TVP models for  their specific application context.
Running the function  \code{shrinkTVP} under  the default model with a fully hierarchical \revised{NG} shrinkage prior with predefined hyperparameters,  estimation of a TVP model becomes as easy  as using the well-known function \code{lm} for a standard linear regression model.
  On the other hand, exploiting numerous advanced options of the package,
  the more experienced user can also explore alternative model specifications such as the Bayesian Lasso or the horseshoe prior and use log-predictive density scores to compare
  various model specifications.}

Various examples of the usage of \pkg{shrinkTVP} were given, and the \code{summary} and \code{plot} methods for
straightforward posterior inference were illustrated.
Furthermore, a predictive exercise with the dataset \code{usmacro.updade} from the package \pkg{bvarsv} was performed, with a focus on the calculation of LPDSs using \code{shrinkTVP}. The default model in \pkg{shrinkTVP} showed better performance than its competitors in terms of cumulative LPDSs.
\comment{While these  examples  were confined to univariate responses,  the package can also be applied in a multivariate context, for instance to the
sparse TVP Cholesky SV model considered in \citet{bit-fru:ach}, exploiting a representation of this model as a system of independent  TVP models
with univariate responses.}

\appendix

\section{Appendix: Full conditional distribution of the latent states}
\label{sec:mccau}
Let $y^\star_t = y_t - \bm x_t \betav$  and  $\Fm_t=\bm x_{t}\Diag{\sqrt{\theta_{1}},\dots,\sqrt{\theta_{d}}}$ for $t=1,\dots,T$. Conditional on all other variables,  the joint density for the state process
   $ \tilde \betav =( \tilde \betav_0, \tilde \betav_1, \dots, \tilde \betav_T)$ is multivariate normal.
This distribution can be written in
   terms of the tri-diagonal precision matrix $\OOmega$  and the mean vector $\cm$ \citep{mcc-etal:sim}:
\begin{eqnarray} \label{postbetav}
    \tilde \betav  | \betav, \QQ ,  \sigma^2_1, \ldots,  \sigma^2_T, y^\star_1, \ldots  y^\star_T
   \sim \Normult{(T+1)  d }{\OOmega^{-1}\cm,\OOmega^{-1} }
\end{eqnarray}
   where:
   \begin{equation*}
\OOmega  =   \begin{bmatrix}
\OOmega_{00}&\OOmega_{01}  & 0 & & \\
\OOmega^{\top}_{01} &  \OOmega_{11}&\OOmega_{12} &0&0&\\
0 &  \OOmega^{\top}_{12}&\OOmega_{22}&\OOmega_{23}&\ddots&\vdots\\
&  0&\OOmega^{\top}_{23}&\ddots&\ddots&0 \\
&  \vdots & \ddots & \ddots &\OOmega_{T-1,T-1}&\OOmega_{T-1,T} \\
&  0&\ldots &0 &\OOmega^{\top}_{T-1,T}&\OOmega_{TT} \\
  \end{bmatrix},  \quad
 \cm  =  \begin{bmatrix}
  \cm_0\\
  \cm_1\\
 \cm_2 \\
  \vdots \\
  \cm_T
  \end{bmatrix} .
\end{equation*}
In this representation, each submatrix $\OOmega_{ts}$ is a matrix of dimension $d \times d$  defined as
\begin{eqnarray*}
 \OOmega_{00} &=&  2 I_d,\\
  \OOmega_{tt} &=& \Fm^{\top}_t   \Fm _t /\sigma^2_t +   2 I_d,\quad t= 1,\dots, T-1,\\
      \OOmega_{TT} &=& \Fm^{\top}_T   \Fm_T /\sigma^2_T  + I_d,\\
   \OOmega_{t-1,t} &=& -  I_d, \quad t=1, \dots, T,
\end{eqnarray*}
where $I_d$ is the $d\times d$ identity matrix and
 $\cm_{t}$ is a  column vector of  dimension $d \times 1$, defined as
\begin{eqnarray*}
\cm_{0}  =  \bfz, \qquad
   \cm_{t}  =  (\Fm^{\top}_t /\sigma^2_t) y_t^{\star},	\quad t = 1,\dots, T .
\end{eqnarray*}
In the homoscedastic case, $\sigma^2_1=\ldots = \sigma^2_T=\sigma^2$.

\section{Multicore LPDS calculation}
\label{sec:multicore}

In the code below, the following \proglang{R} packages are used:
\pkg{doParallel} \citep{wes:doP}, \pkg{foreach} \citep{wes:for},
\pkg{zoo} \citep{zei:zoo}, and \pkg{RhpcBLASctl} \citep{Nak:Rhp}.

\begin{CodeInput}
R> # Calculate LPDS in multicore
R> # Load libraries for multicore computations
R> library("doParallel")
R> library("foreach")
R> 
R> # For manipulating dates
R> library("zoo")
R> 
R> # Load library for controlling number of BLAS threads
R> library("RhpcBLASctl")
R> 
R> # Define how many periods to calculate LPDS for
R> Tmax <- nrow(us_data) - 1
R> T0 <- Tmax - 49
R> 
R> # Determine number of cores to be used and register parallel backend
R> ncores <- 4
R> cl <- makeCluster(ncores)
R> registerDoParallel(cl)
R> 
R> lpds <- foreach(t = T0:Tmax, .combine = "cbind",
+    .packages = c("RhpcBLASctl", "shrinkTVP"),
+    .errorhandling = "pass") %dopar% {
+  
+    set.seed(t)
+  
+    niter <- 30000
+    nburn <- 15000
+    nthin <- 5
+  
+    # Set number of BLAS threads, so they dont interfere with each other
+    blas_set_num_threads(1)
+  
+    # Create data_t from all data up to time t and
+    # y_test and x_test from data at time t+1
+    data_test <- us_data[t+1,]
+    data_t <- us_data[1:t,]
+  
+    # Run MCMC to calculate all LPDS
+    # Fully hierarchical triple gamma
+    res_FH_TG <- shrinkTVP(inf ~ inf_lag + une_lag + tbi_lag, data = data_t,
+      mod_type = "triple", niter = niter, nburn = nburn, nthin = nthin)
+  
+    # Hierarchical triple gamma
+    res_H_TG <- shrinkTVP(inf ~ inf_lag + une_lag + tbi_lag, data = data_t,
+      mod_type = "triple", niter = niter, nburn = nburn, nthin = nthin,
+      learn_a_xi = FALSE, learn_a_tau = FALSE,
+      learn_c_xi = FALSE, learn_c_tau = FALSE)
+  
+    # Non-hierarchical triple gamma
+    res_TG <- shrinkTVP(inf ~ inf_lag + une_lag + tbi_lag, data = data_t,
+      mod_type = "triple", niter = niter, nburn = nburn, nthin = nthin,
+      learn_kappa2_B = FALSE, learn_lambda2_B = FALSE,
+      learn_a_xi = FALSE, learn_a_tau = FALSE,
+      learn_c_xi = FALSE, learn_c_tau = FALSE)
+  
+    # Hierarchical horseshoe
+    res_H_HS <- shrinkTVP(inf ~ inf_lag + une_lag + tbi_lag, data = data_t,
+      mod_type = "triple", niter = niter, nburn = nburn, nthin = nthin,
+      learn_a_xi = FALSE, learn_a_tau = FALSE,
+      learn_c_xi = FALSE, learn_c_tau = FALSE,
+      a_xi = 0.5, a_tau = 0.5, c_xi = 0.5, c_tau = 0.5)
+  
+    # Non-hierarchical horseshoe
+    res_HS <- shrinkTVP(inf ~ inf_lag + une_lag + tbi_lag, data = data_t,
+      mod_type = "triple", niter = niter, nburn = nburn, nthin = nthin,
+      learn_kappa2_B = FALSE, learn_lambda2_B = FALSE,
+      learn_a_xi = FALSE, learn_a_tau = FALSE,
+      learn_c_xi = FALSE, learn_c_tau = FALSE,
+      a_xi = 0.5, a_tau = 0.5, c_xi = 0.5, c_tau = 0.5)
+  
+    # Fully hierarchical double gamma
+    res_FH_DG <- shrinkTVP(inf ~ inf_lag + une_lag + tbi_lag, data = data_t,
+      niter = niter, nburn = nburn, nthin = nthin,
+      hyperprior_param = list(nu_tau = 1, nu_xi = 1))
+  
+    # Hierarchical double gamma
+    res_H_DG <- shrinkTVP(inf ~ inf_lag + une_lag + tbi_lag, data = data_t,
+      niter = niter, nburn = nburn, nthin = nthin,
+      learn_a_xi = FALSE, learn_a_tau = FALSE,
+      a_xi = 0.1, a_tau = 0.1)
+  
+    # Non-hierarchical double gamma
+    res_DG <- shrinkTVP(inf ~ inf_lag + une_lag + tbi_lag, data = data_t,
+      niter = niter, nburn = nburn, nthin = nthin,
+      learn_a_xi = FALSE, learn_a_tau = FALSE,
+      a_xi = 0.1, a_tau = 0.1,
+      learn_kappa2_B = FALSE, learn_lambda2_B = FALSE)
+  
+    # Hierarchical Lasso
+    res_H_LS <- shrinkTVP(inf ~ inf_lag + une_lag + tbi_lag, data = data_t,
+      niter = niter, nburn = nburn, nthin = nthin,
+      learn_a_xi = FALSE, learn_a_tau = FALSE,
+      a_xi = 1, a_tau = 1)
+  
+    # Non-hierarchical Lasso
+    res_LS <- shrinkTVP(inf ~ inf_lag + une_lag + tbi_lag, data = data_t,
+      niter = niter, nburn = nburn, nthin = nthin,
+      learn_a_xi = FALSE, learn_a_tau = FALSE,
+      a_xi = 1, a_tau = 1,
+      learn_kappa2_B = FALSE, learn_lambda2_B = FALSE)
+  
+    # Ridge regression
+    res_FV <- shrinkTVP(inf ~ inf_lag + une_lag + tbi_lag, data = data_t,
+      mod_type = "ridge", niter = niter, nburn = nburn, nthin = nthin)
+  
+  
+    lpds_res <- c(LPDS(res_FH_TG, data_test),
+      LPDS(res_H_TG, data_test),
+      LPDS(res_TG, data_test),
+      LPDS(res_H_HS, data_test),
+      LPDS(res_HS, data_test),
+      LPDS(res_FH_DG, data_test),
+      LPDS(res_H_DG, data_test),
+      LPDS(res_DG, data_test),
+      LPDS(res_H_LS, data_test),
+      LPDS(res_LS, data_test),
+      LPDS(res_FV, data_test))
+  
+    rm(list = ls()[!ls() %in% c("lpds_res", "us_data")])
+  
+    return(lpds_res)
+  }
R> stopCluster(cl)
R> 
R> 
R> cumu_lpds <- apply(lpds, 1, cumsum)
R> color <- c(rep("cyan3", 3),
+             rep("firebrick3", 2),
+             rep("forestgreen", 3),
+             rep("yellow2", 2),
+             "black")
R> lty <- c(1:3, 1:2, 1:3, 1:2, 1)
R> 
R> # Plot results
R> par(mar=c(6,4,1,1))
R> colnames(cumu_lpds) <- c("Fully hierarchical NGG",
+    "Hierarchical NGG",
+    "NGG",
+    "Hierarchical Horseshoe",
+    "Horseshoe",
+    "Fully hierarchical NG",
+    "Hierarchical NG",
+    "NG",
+    "Hierarchical Lasso",
+    "Lasso",
+    "Ridge Regression")
R> 
R> matplot(cumu_lpds, type = "l", ylab = "Cumulative LPDS",
+    xaxt = "n", xlab = "", col = color, lty = lty, lwd = 1.5)
R> 
R> # Extract labels from time series
R> labs = as.yearmon(time(usmacro.update))[T0:Tmax + 1][c(FALSE, TRUE)]
R> 
R> # Create custom axis labels
R> axis(1, at = (1:length(T0:Tmax))[c(FALSE, TRUE)], labels = FALSE)
R> text(x=(1:length(T0:Tmax))[c(FALSE, TRUE)],
+    y=par()$usr[3]-0.05*(par()$usr[4]-par()$usr[3]),
+    labels=labs, srt=45, adj=1, xpd=TRUE)
R> 
R> # Add legend
R> legend("topright", colnames(cumu_lpds), col = color,
+    lty = lty,lwd = 1.5, bty = "n", cex = 0.8)
\end{CodeInput}

\bibliography{sylvia_kyoto_2}

\end{document}